\documentstyle[12pt,preprint,prd,aps,subeqn]{revtex}
\begin{document}

\include{psfig}
\include{johnmacro}
\def\pp{\hbox{pp}\tcar}

\title{Gallium Solar Neutrino Experiments: Absorption Cross Sections,
Neutrino Spectra,  and
Predicted Event Rates}

\author{John N. Bahcall }
\address{School of Natural Sciences, Institute for Advanced 
Study\\
Princeton, NJ 08540\\}
\bigskip

\maketitle
\begin{abstract}
\small\tighten
Neutrino absorption cross sections for ${\rm^{71}Ga}$
are calculated for all solar neutrino sources
with standard energy spectra, and for  laboratory sources of ${\rm^{51}Cr}$
and ${\rm^{37}Ar}$;  
the calculations  include, where appropriate,  
the thermal energy of  decaying
solar ions and 
use improved nuclear and atomic data.
The ratio, $R$, of measured (in GALLEX and SAGE) to calculated ${\rm ^{51}Cr}$ capture rate
is $R = 0.95 \pm 0.07~{\rm (exp)}~+~^{+0.04}_{-0.03}~{\rm (theory)}$.
Cross sections are also calculated   for  specific
neutrino energies chosen so that a spline fit 
determines accurately 
the event rates in a gallium detector even if new physics
changes the energy spectrum of solar neutrinos. Theoretical
uncertainties are estimated for cross sections at specific energies
and for standard neutrino energy spectra.
Standard 
energy spectra are presented for $pp$ and CNO neutrino
sources in the appendices.  
Neutrino fluxes predicted by standard solar models, 
corrected for
diffusion,  have been in the 
range $120$ SNU to $141$ SNU since 1968.
\end{abstract}

\section{Introduction}
\label{intro}
Gallium solar neutrino experiments are, at present, the only detectors
capable of detecting the 
fundamental $pp$ neutrinos,  which constitute about 90\% 
of the neutrinos predicted by standard solar models to come from the sun.
The pioneering gallium solar neutrino 
experiments, GALLEX\cite{gallexresults}
and SAGE\cite{sageresults}, are also unique in having been directly
tested for efficiency of neutrino detection  with a radioactive source,
${\rm^{51}Cr}$\cite{gallexcr1,gallexcr2,cribier88,sagecr}.  
Moreover, the good agreement between the results of the  
two independent 
experiments, one of which uses  gallium in chloride solution 
(GALLEX) and the other in a metallic form (SAGE),
 has led to increased confidence in the measured event
rates. 
The results of the gallium experiments provide fundamental constraints
on solar models and challenge the prediction  of minimal electroweak
theory that essentially nothing happens to neutrinos after they are 
created in the center of the sun.  

The Gallium Neutrino Observatory (GNO) collaboration\cite{gno} has 
recently been formed  to measure 
the solar
neutrino event rate in a gallium detector
over many years (at least one solar cycle)
and with increased precision.
The experiment, which may ultimately involve 100 tons of gallium,
 is designed to reduce both the systematic and the
statistical errors so that an accuracy of about $5$\%, or $4$ SNU,
will be achieved if the final best-estimate event rate is $80$ SNU.
\footnote{A SNU is a convenient product of flux times cross section
first defined in footnote 10 of Ref. \cite{bahcall69}, in 1969, to 
be $10^{-36} {\rm ~ interactions~per~target~atom~per~sec}$.}

Motivated by  the great importance of gallium solar
neutrino experiments and the improvements possible with GNO,
my goal in  this paper is 
to calculate as accurately as
possible the cross sections for absorption of solar neutrinos in a gallium
detector and to explore more broadly the constraints on solar nuclear
fusion provided by existing and future gallium experiments.
The theoretical uncertainties in the capture cross sections, which are
a function of neutrino energy, limit the ultimate interpretation of
the observed results.
Therefore, I devote a large part of the present paper
to evaluating quantitatively 
the uncertainties that exist in the
calculations of neutrino absorption cross sections. 

I begin by summarizing in Sec.~\ref{input} the most important
experimental data regarding the ${\rm^{71}Ga}$-${\rm^{71}Ge}$ system.
I describe in Sec.~\ref{smalltheory} how I evaluate the atomic
effects of electron exchange and imperfect overlap between initial and
final eigenstates, as well as the forbidden nuclear beta-decay
corrections.  I make use of new Dirac-Fock calculations of the
electron density at the nucleus in a ${\rm ^{71}Ge}$ atom, in order to
evaluate more accurately than was previously possible the $f$-value
for ${\rm ^{71}Ge}$ electron capture.
In Sec.~\ref{crsection}, I calculate the cross section for
the absorption by ${\rm^{71}Ga}$ of neutrinos from  ${\rm^{51}Cr}$ beta decay and
compare the calculated value with 
the value inferred by the GALLEX and SAGE experiments and with the
previous calculation.
I describe in Sec.~\ref{exciteduncertainties} the procedure I
use to evaluate the uncertainties
due to excited state transitions.
 I make conservative assumptions about the BGT values
that are 
determined by $(p,n)$ reactions  and follow Anselmann {\it et
al.}\cite{gallexcr1} and Hata and
Haxton\cite{hata95} in using the results of the ${\rm^{51}Cr}$ experiments
performed by GALLEX and SAGE to constrain the neutrino absorption
cross sections for  transitions from the ground state of ${\rm^{71}Ga}$ to
the lowest two excited states of ${\rm^{71}Ge}$ for which allowed captures
are possible.   I describe in Sec.~\ref{thermaleffects}
the contributions to the energy spectra from the thermal 
energy of the fusing particles that
produce neutrinos; these thermal energy contributions are included here for the first
time in the calculation of the absorption cross sections.

If standard solar models and the minimal standard electroweak theory
are correct, then
the $pp$ neutrinos provide the largest predicted contribution to
gallium solar neutrino experiments. In Sec.~\ref{ppsection}, I
evaluate the absorption cross section for $pp$ neutrinos including
for the first time the effect of the thermal energy of the fusing
protons.  I present 
in Sec.~\ref{continuumcross} the results of
calculations of the cross sections for the beta-decaying sources 
${\rm^8B}$, ${\rm^{13}N}$, ${\rm^{15}O}$, and ${\rm^{17}F}$,
emphasizing
the uncertainties caused by transitions to excited states in
${\rm^{71}Ge}$.  
I present the cross section for the highest energy solar neutrinos,
the $hep$ neutrinos, in Sec.~\ref{hepcontinuum}.

The average 
cross section for the absorption of the ${\rm^7Be}$ neutrino lines is 
important in understanding the implications of gallium solar neutrino
measurements. I calculate in Sec.~\ref{be7pluspep} the average 
cross sections for the 
${\rm^7Be}$ neutrino lines and for the $pep$ line, including the thermal
energy of the solar electrons and ions.  
I also calculate in Sec.~\ref{be7pluspep} the cross section for 
absorption of neutrinos from a laboratory source of ${\rm^{37}Ar}$
neutrinos. 

Particle physics explanations
of the solar neutrino measurements generally result in a modified
neutrino energy spectrum for the electron-type neutrinos.  
Therefore, in
Sec.~\ref{crossspecific} I present  calculated best-estimate 
cross sections, and $3\sigma$ different cross sections, 
for a representative set
of specific neutrino energies. 

In Sec.~\ref{eventrates}, I calculate the 
event rate predicted by the 
current best standard solar model
and compare the results with the rates
measured by GALLEX and SAGE.  
I show in a figure the rates predicted by all standard solar models
calculated by collaborators and myself since 1963.  
I also determine the rates predicted by
solar models with crucial nuclear reactions artificially set equal to
zero.
I summarize and
discuss the main results in Sec.~\ref{summary}.

This paper also presents some additional data that are of general use
in stellar evolution studies or for solar neutrino investigations.
The average energy loss for each neutrino energy source, which is
important for stellar evolution calculations, is given in the text
that discusses the absorption cross section for that particular source.
I tabulate in Appendix~A the $pp$ solar neutrino energy spectrum and in
Appendix~B the energy spectra for the CNO neutrino sources.

Unless stated otherwise, all nuclear data (including  lifetimes,
branching ratios, and mass differences, as well as their associated
uncertainties),  and also atomic binding
energies, are  taken from the 1996
8th edition of the Table of Isotopes\cite{firestone96}.

\section{Gallium--Germanium  Data}
\label{input}

I summarize in this section the basic data for the gallium--germanium
system that are needed for the calculation of solar neutrino cross
sections.  I begin by listing in Sec.~\ref{measuredgage}
the best values and uncertainties for
the most important measured atomic and nuclear quantities. Next,  I
calculate in Sec.~\ref{secsigmazero} the characteristic dimensional
cross section factor, $\sigma_0$, for gallium--germanium transitions.
I conclude  by discussing in Sec.~\ref{excited} what is known
from $(p,n)$ measurements about the matrix elements for transitions 
from the ground state of ${\rm^{71}Ga}$ to different excited states of 
${\rm^{71}Ge}$.

Figure~\ref{gagefigure} illustrates
the most important neutrino transitions
for a gallium solar neutrino experiment, provided that the incident
neutrino flux is dominated, as expected on the basis of standard solar
models, by neutrinos with energies less than 1~MeV.

\subsection{Measured Nuclear and Atomic Properties}
\label{measuredgage}

The principal input data needed for the calculation of the neutrino
cross sections are the atomic number, $Z = 31$ and the atomic mass,
$A = 71$,
of ${\rm^{71}Ga}$, 
the measured
electron capture lifetime of ${\rm^{71}Ge}$, $\tau_{1/2}$,  the
neutrino energy threshold, $E_{\rm Threshold}$, and the electronic
binding energies in ${\rm^{71}Ga}$.

The  lifetime for ${\rm^{71}Ge}$ electron capture 
has been measured accurately by Hampel and Remsberg and is\cite{hampel85}
\begin{equation}
\tau_{1/2} = 11.43 \pm 0.03\  d
\label{galifetime}
\end{equation}

The energy threshold has been measured in a number of different
experiments. The value of $E_{\rm Threshold} = 233.2 \pm 0.5$ keV 
due to Hampel and Schlotz\cite{hampel84} was used in the previous
calculations\cite{BU88,bahcall89} of gallium absorption cross sections.
There have subsequently been three additional measurements in the
context of the search for a possible $17$ keV neutrino; they are:
$229.1 \pm 0.6$ keV\cite{zlimen91}, $232.1 \pm 0.1$ keV
\cite{diGregorio93}, and $232.65^{+0.17}_{-0.12}$\cite{lee95}.
The result from the 
first of these three measurements is somewhat uncertain since these
authors\cite{zlimen91} found evidence for a $17$ keV neutrino in their
internal bremsstrahlung spectrum. A weighted 
average of all the available measurements, including estimates of
systematic errors, has been computed by Audi and Wapstra\cite{audi97},
who find
\begin{equation}
E_{\rm Threshold} = 232.69 \pm 0.15\,\,{\rm keV} .
\label{gathreshold}
\end{equation}
The s-shell binding energies of the K, L, and M electrons in
${\rm^{71}Ga}$, 
which are
needed below for  the calculation of $\sigma_0$, are, respectively,
$10.37$ keV, $1.30$ keV, and $0.16$ keV.

\subsection{Calculation of $\sigma_0$}
\label{secsigmazero}

Neutrino absorption cross sections from the ground state of ${\rm^{71}Ga}$ to 
the ground state of ${\rm^{71}Ge}$ are inversely proportional to the $ft$-value
for the inverse process, the 
electron capture decay of ${\rm^{71}Ge}$ to ${\rm^{71}Ga}$. 
The precise form of this relation is given in Eq.~(11) of 
 Ref. \cite{bahcall78}, which defines a characteristic 
neutrino absorption cross section,
$\sigma_0$, in terms of the electron capture rate from the ground
state of the daughter nucleus
produced by neutrino capture.
The quantity $\sigma_0$ is used as an overall scale factor in detailed
numerical calculations of neutrino absorption cross sections.

Inserting the best available estimates for the 
 electron capture lifetime, Eq.~(\ref{galifetime}), and 
energy threshold, Eq.~(\ref{gathreshold}), in
Eq.~(11) of Ref. \cite{bahcall78}
yields
\begin{equation}
\sigma_0\ =\ \frac{1.2429 \times 10^{-47}\ {\rm cm^2}}{\Sigma_i q^2_i g^2_i}~,
\protect\label{defsigmazero}
\end{equation}
where $q_i = E_{\rm threshold} - E_{{\rm binding},i}$ and $g^2_i$ is the
square of the radial electron wave function,
averaged over the nuclear volume, for the $i^{\rm th}$
atomic state in ${\rm ^{71}Ge}$ .  The
numerical coefficient in Eq.~(\ref{defsigmazero}) is appropriate when
$q_i$ is measured in $m_e c^2$ and $g^2_i$ is measured in $(\hbar/m_e
c)^{-3}$.

I have recalculated $\sigma_0$ using the results of a special series
of evaluations of the ${\rm ^{71}Ge}$ electron wave functions
generously performed for application to this work by I. Grant
\cite{grant}, W. Johnson \cite{johnson}, and M. Chen \cite{chen}.  All
three of the calculations use relativistic, self-consistent
Dirac-Fock codes that include the effects of
 finite sized nuclei, the Breit interaction, vacuum polarization, and
self-energy corrections.  Details of the codes used
are supplied in the references just cited.  I have averaged the values of
$g^2_i$ over the nuclear volume using data supplied by Grant, Johnson,
and Chen.
The total spread among the three
calculations is only $\pm 0.2\%$, which does not contribute
significantly to the overall error budget of the cross section calculations.

Using the new relativistic Hartree Fock
calculations~\cite{grant,johnson,chen} and 
the slightly improved energy threshold given in
Eq.~(\ref{gathreshold}),
I find

\begin{equation}
\sigma_o = 8.611 \times 10^{-46}~{\rm cm}^2 ,
\label{sigmazero}
\end{equation}
with an overall (effective $3\sigma$) systematic 
uncertainty of $0.4$\% that is
common to all cross sections quoted in this paper and which should be
treated as an additional  theoretical uncertainty (not included
elsewhere) in precise error analyses of gallium solar neutrino
experiments. The 
 value of $\sigma_0$ given in Eq.~(\ref{sigmazero})
is about $0.5$\%  less than the value of 
$\sigma_0 = 8.8012\times 10^{-46}~{\rm cm}^2 $ that I have used since
1984 (see Refs.~\cite{BU88,bahcall89,krofcheck85}).  All the published
calculations on the implications of 
gallium solar neutrino experiments with which I am
familiar have also 
made use of this previous determination of $\sigma_0$.

The short-distance high momentum loop radiative corrections are
automatically taken into account by scaling all of the neutrino
capture cross sections relative to the electron capture rate that 
determines $\sigma_0$\cite{sirlin,marciano}. The additional radiative
corrections are expected to be smaller than the short distance
corrections and therefore significantly less 
than other uncertainties estimated in
this paper\cite{marciano}.

\subsection{Transitions to Excited States}
\label{excited}

The most important   uncertainties in the calculation of  absorption
cross sections for solar neutrinos incident on ${\rm^{71}Ga}$ are 
related, for all but the 
lowest energy neutrinos, to the transition matrix elements to excited 
states in ${\rm^{71}Ge}$ (see, e.g., Refs. \cite{BU88,bahcall78,hata95}).
In what follows, I shall make use of the BGT values for transitions
to excited states of ${\rm^{71}Ge}$ that are estimated by 
studying $(p,n)$ reactions.
The BGT values determined by Krofcheck and his 
colleagues\cite{krofcheck85,krofcheck87}
are listed in Table~\ref{BGT}. In order to help 
make clear which transitions 
are most important, I have presented the estimated BGT values
relative to the BGT value  of the ground state to ground state transition.
The resolution obtained in the experiments of Krofcheck {\it et al.} is about 
300 keV; this resolution is not always sufficient to identify 
the particular
final state(s) in ${\rm^{71}Ge}$ to which the strength of the $(p,n)$ reaction 
refers (cf. the discussion of the cross section for the $pep$ line in
Sec.~\ref{pepline}).
The transition to the $5/2^-$ first excited state of ${\rm^{71}Ge}$ at 
an excitation energy of $175$ keV was too weak to be measured and only
an upper limit was determined. 
In the discussion that follows, I shall use for
definiteness a BGT value that is one half of the measured upper
limit, i.e., 
${\rm BGT(175~keV)/(BGT)}_{\rm g.s.} = 0.028$.

I follow Anselmann {\it et al.}\cite{gallexcr1} and 
Hata and Haxton\cite{hata95} in using the
GALLEX\cite{gallexcr1,gallexcr2,cribier88} and SAGE\cite{sagecr}
measurements of ${\rm^{51}Cr}$ neutrino absorption by ${\rm^{71}Ga}$ to
constrain the BGT values for solar neutrino detection (see
Sec.~\ref{crcrossconstraint}).
 
\section{Overlap, Exchange, and Forbidden Effects}
\label{smalltheory}

I summarize in this section how I calculate atomic overlap and
exchange effects and nuclear forbidden effects.

\subsection{Overlap and Exchange Effects}
\label{overlapandexchange}

The change in nuclear charge by one unit in beta decay and in 
neutrino capture causes the initial
and final atomic eigenstates to overlap imperfectly, which is known as 
the ``overlap'' effect in atomic beta decay.  Antisymmetrization
between bound and continuum electrons, the ``exchange effect'',
has a measurable effect in determining
electron capture ratios and 
decreases the calculated cross sections 
for  neutrino capture reactions by a small, calculable  factor.

Overlap and exchange effects were first  discussed in detail in
Ref. \cite{bahcall63} and applied in Ref. \cite{bahcall78} to 
the calculation of neutrino capture cross sections.  
Experiments on electron capture ratios provide strong
evidence for the validity of the electron exchange corrections
(see Ref.~\cite{bahcall65} for an early discussion).
I use  
Eq.~(13) and Eq.~(14) of Ref.~\cite{bahcall78} to evaluate numerically
the atomic overlap and exchange effects.
In previous calculations, I have been content with showing that these
corrections are small, less than or of order 1\%\cite{bahcall78}, and
have not included them explicitly in the numerical calculations.

The imperfect overlap between initial and final atomic
states adds
$0.09$ keV 
to the usually-tabulated mass difference between neutral
atoms for the case of neutrino capture by gallium; this small 
energy difference, which slightly increases the threshold energy, is 
not included in Eq.~(\ref{gathreshold}), but is taken account
of in the numerical calculations described in the present paper.

The overlap effect is not significant for our purposes. Even
for the low-energy $pp$ neutrinos, the overlap effect decreases the
calculated absorption cross section by less than $0.1$\%.

Exchange effects between the final continuum electron and the
electrons bound in the initial atom interfere in a way that reduces
slightly the calculated capture rate\cite{bahcall63,bahcall78}. 
In the calculations described later in this paper, I 
evaluate the exchange effect for  gallium cross sections by using 
Eq.~(14)of Ref. \cite{bahcall78}.
For the $pp$ cross section which is evaluated  in 
Sec.~\ref{ppsection}, 
exchange effects reduce the calculated 
cross section by $0.4$\%.  In all other
cases, the calculated 
effects of electron exchange are negligible, because the
characteristic energies of solar neutrinos (MeV) are much larger than
the characteristic binding energies of the atomic electrons (keV).

\subsection{Forbidden Corrections}
\label{forbidden}

Forbidden corrections to nuclear beta-decay have been calculated by
many workers.  I follow here the prescription due to Holstein and 
Treiman\cite{holstein71}, which has been applied 
by Bahcall and Holstein\cite{holstein86} to solar neutrino
problems.  I use for the best-estimate calculations presented in 
the present paper the 
approximations that are 
described in the Appendix of Ref.~\cite{holstein86}.

Since the forbidden corrections involve estimates of 
unmeasurable nuclear matrix
elements, I regard the calculated  corrections as only an
indication of the likely size of forbidden effects.  
With the Bahcall-Holstein approximations, the forbidden effects vary
slowly between 2\% and 2.5\% at neutrino energies less than about 10
MeV and then change sign becoming approximately zero at $15$ MeV. 
This cancellation near $15$ MeV is presumably an artifact of the
choice by Bahcall and Holstein of characteristic nuclear parameters
that were intended to apply approximately over a wide range of nuclei.
In
the energy regime relevant to supernovae neutrinos but 
just beyond the reach of nearly all solar neutrinos, $15$ MeV to 
$30$ MeV, the calculated 
forbidden corrections rise rapidly, approximately proportional
to the square of the recoil electron energy. 

For the purpose of estimating uncertainties, I adopt the conservative
approach of replacing the decrease in the calculated forbidden
corrections in the region $10$ MeV to $15$ MeV by 
a monotonic estimate of $0.025(q/10~{\rm MeV})^2$ for neutrino
energies above $10$ MeV. This procedure ignores the cancellation that 
occurs near $15$ MeV, but otherwise gives a relatively accurate
numerical
representation of the higher-energy  forbidden corrections as
described in Ref.~\cite{holstein86}.

I adopt three
times the best-estimate forbidden correction as a 
$3\sigma$ uncertainty due to forbidden corrections. 
I ignore the calculated sign of the forbidden corrections and instead
assume that both the estimated minimum and the estimated maximum cross
sections have a contribution from forbidden terms 
of equal magnitude (added
quadratically with other uncertainties).

Fortunately, for solar neutrino cross sections, the 
forbidden corrections are never fractionally very large. For all
cases except for the $pp$ neutrinos, 
the uncertainties due to transitions to excited
states are much larger than the uncertainties due to forbidden
corrections. However the uncertainties become very large, about
a factor of three,  above 
$25$ MeV (see Sec.~\ref{crossspecific}).

\section{The Chromium Neutrino Absorption Cross Section}
\label{crsection}

The overall efficiency for the detection of neutrinos with 
radiochemical gallium detectors has been
measured  directly in two historic experiments 
by the GALLEX\cite{gallexcr1,gallexcr2,cribier88} and the 
SAGE\cite{sagecr} solar neutrino collaborations using intense ${\rm^{51}Cr}$
sources of low energy neutrinos. Originally proposed by 
Kuzmin\cite{kuzmin67}
and Raghavan\cite{raghavan78}, these calibration experiments show that the
neutrino 
detectors work as expected.  The ${\rm^{51}Cr}$ neutrino sources are especially 
useful for testing the detection efficiency since the chromium
neutrinos are  similar in energies to the $pp$ and ${\rm^7Be}$ solar 
neutrinos to which the gallium detectors are most sensitive.
In addition to being the first direct tests of solar neutrino experiments
with an artificial source of neutrinos, 
the calibration 
results also improve by more than an order of magnitude previous limits
on $\Delta m^2$ for electron-neutrino oscillation experiments at 
accelerators\cite{bkl95}.
Moreover, Anselmann {\it et al.}\cite{gallexcr1} and 
Hata and Haxton\cite{hata95} have shown 
that the test results provide
useful direct constraints on the BGT values for excited state transitions from 
the ground state of ${\rm^{71}Ga}$ to excited states of ${\rm^{71}Ge}$ that must 
otherwise be inferred from the 
less-easily interpreted $(p,n)$ experiments
(cf. Sec.~\ref{input}).

Figure~\ref{crdecayscheme} shows the 
four neutrino lines that are produced by the 
decay of ${\rm^{51}Cr}$\cite{firestone96}. 

\subsection{Experimental Results}
\label{expcr}

The experimental results 
on ${\rm^{51}Cr}$ neutrino absorption 
are reported\cite{gallexcr1,gallexcr2,sagecr} as
a ratio, $R$, of the measured cross section 
to the value of the cross section
calculated by Bahcall and Ulrich in 1988\cite{BU88}.  Thus

\begin{equation}
R \equiv {\sigma {\rm (^{51}Cr)_{measured}}  \over \sigma {\rm
(^{51}Cr)_{BU88}}} .  
\label{Rdef}
\end{equation}
The value of the previous standard cross section is\cite{BU88}
\begin{equation}
\sigma {\rm (^{51}Cr)_{BU88}} = 59.2 \times 10^{-46} 
(1 \pm 0.1)~{\rm cm^{-2}},\
\ 3\sigma ,
\label{BU88value}
\end{equation}
where the total 
theoretical error quoted in Eq.~(\ref{BU88value}) is an effective
$3\sigma$ uncertainty. 

The measured values, with their quoted $1\sigma$ errors
are\cite{gallexcr1,gallexcr2} 
\begin{equation}
R {\rm (GALLEX)} = 0.92 \pm 0.08,\ \ 1\sigma
\label{Rgallex}
\end{equation}
and\cite{sagecr}
\begin{equation}
R {\rm (SAGE)} = 0.95 \pm 0.12,\ \ 1\sigma  .
\label{Rsage}
\end{equation}
The weighted average value for $R$, $\langle R\rangle$, 
obtained by  combining the results from the two 
experiments, is very similar to the GALLEX value.  One finds:
\begin{equation}
\langle R\rangle = 0.93 \pm 0.07,\ \ 1\sigma .
\label{Rav}
\end{equation}
The errors quoted  in 
Eq.~(\ref{Rgallex})--Eq.~(\ref{Rav}) 
for the experimental results
do not include the uncertainty in the 
theoretical calculation that is given in Eq.~(\ref{BU88value}).

\subsection{Measured Quantities Characterizing Chromium Decay}

The Q-value or atomic mass difference between the 
ground states of ${\rm^{51}Cr}$
and ${\rm^{51}V}$ is

\begin{equation}
Q = 752.73 \pm 0.24\ {\rm keV} .
\label{Qvaluecr}
\end{equation}
A  preliminary result from the remeasurement of the Q-value 
using internal bremstrahlung from the ${\rm^{51}Cr}$
decay carried out by Hampel, Hartmann, and Heusser\cite{hampel97} gives a result
in good agreement with Eq.~(\ref{Qvaluecr}). 

As shown in Fig.~\ref{crdecayscheme}, 
electron capture by ${\rm^{51}Cr}$ leads to the ground state of ${\rm^{51}V}$ 
with a branching ratio of 90.12\% and to the first excited state with
a branching ratio of 9.88\% . The neutrino energy released
in the electron capture reaction leading to the ${\rm^{51}Cr}$ ground state is
$Q - E(K) = 747.27$ keV, where $E(K) = 5.46$ keV is the 
binding energy of the K-electron
in ${\rm^{51}V}$. The neutrino energy corresponding to L-capture is 
$Q = 752.73- 0.63$ 
keV or $752.10$ keV.  The neutrino energies that result from K and L 
captures 
to the first excited state 
of ${\rm^{51}V}$ are $432.02$ keV  and $427.19$ keV, respectively.
The measured capture ratio is\cite{heuer66} 
\begin{equation}
L/K = 0.104 \pm 0.003,\ \ 1\sigma .
\label{KoverL}
\end{equation}

\subsection{Best Estimate}
\label{crbestestimate}

The best-estimate neutrino absorption cross section 
averaged over the four neutrino lines of ${\rm^{51}Cr}$ is
\begin{equation}
\sigma {\rm (^{51}Cr)_{\rm Best}} = 58.1 (1^{+0.036}_{-0.028} )
\times 10^{-46}~{\rm
cm^2},\ \ 1\sigma .
\label{sig51best}
\end{equation}
In computing the cross section given in Eq.~(\ref{sig51best}), I have
used the Q-value for the chromium decay given in 
Eq.~(\ref{Qvaluecr}), the  
gallium threshold given in Eq.~(\ref{gathreshold}), and 
the characteristic gallium
cross section, $\sigma_0$, given in Eq.~(\ref{sigmazero}). 
The current best theoretical estimate for the average cross section
for ${\rm^{51}Cr}$
neutrinos given in Eq.~(\ref{sig51best}) 
is 2\% smaller  than the previous value (cf. Eq.~(\ref{BU88value})~).
The value computed here is 5.4\% larger (less than $1 \sigma$ larger)
than the best-estimate experimental value of $55.1 \times 10^{-46}~{\rm
cm}^2$ [cf. Eq.~(\ref{Rav})].

Most of the computed cross section comes from transitions between the 
ground state of ${\rm^{71}Ga}$ and the ground state of ${\rm^{71}Ge}$.  The
fraction of the total cross section that arises from ground state to
ground state transitions is 
\begin{equation}
\sigma {\rm (^{51}Cr)_{\rm gs}}/\sigma {\rm (^{51}Cr)_{\rm Best}} = 
0.95 . 
\label{sig51gs}
\end{equation}

\subsection{Uncertainties}
\label{cruncertainties}

The largest uncertainty in the prediction of the ${\rm^{51}Cr}$ absorption
cross section arises from the poorly known matrix elements for the 
transitions from the ground state of ${\rm^{71}Ga}$ to the excited states
of ${\rm^{71}Ge}$\cite{BU88,bahcall78,hata95} .  I have used as
best-estimates for excited state transitions the BGT values that
were determined from $(p,n)$ measurements by Krofcheck and his
associates\cite{krofcheck85,krofcheck87}
(see Table~\ref{BGT} ).
I have taken  the minimum contribution from excited states to be zero and
regard this decrease, $-4.8$\%,  from the best estimate value 
as a  $3\sigma$ change.

For the maximum $3\sigma$ contribution 
from excited states, I have multiplied 
the BGT values determined by $(p,n)$ measurements by 
a factor of two.  More explicitly, I multiplied the measured 
$(p,n)$ upper limit
BGT value to the first excited state, ${\rm BGT/BGT}_{\rm g.s.} < 0.056$, by
a factor of two and the measured value to the second excited state,
${\rm BGT/BGT}_{\rm g.s.} = 0.146$, also by a factor of two.  In all ten
cases in which $(p,n)$-inferred BGT values for weak transitions 
 have been compared to
accurately measured beta-decay matrix elements, the $(p,n)$ values are
about equal to  or much larger than the true beta-decay matrix
elements. It is possible\cite{hata95}  that this consistent
trend is due to a special selection rule operating in all 10 of the
cases in which both the beta-decay and the $(p,n)$ measurements have
been made accurately.  However, there is additional information
available to support the procedure adopted here.
The two cases which are most relevant to the current
discussion occur in the decay of ${\rm^{37}Ca}$ to ${\rm^{37}K}$; these are the
only two cases with which I am familiar in which the corresponding 
beta-decay matrix elements are very
small, comparable to the BGT values determined by the $(p,n)$
measurements for the weak transitions to the first two excited states
of ${\rm^{71}Ge}$. For these two weak transitions, the measured BGT values
from beta-decay are between one and two orders of magnitude smaller
than the BGT values inferred from $(p,n)$ measurements\cite{hata95}.  
Therefore, the 
upper limit change, $+8.4$\%, 
determined by multiplying the $(p,n)$-inferred BGT value, and the upper
limit BGT value to the lowest excited state,  both by a factor of two 
seems like a reasonable effective $3\sigma$
upper limit to the contribution of excited state transitions to the
best-estimate value. 

The next largest uncertainty is from forbidden corrections to the
beta-decay matrix elements\cite{holstein86}. Omitting entirely the
forbidden corrections, decreases the calculated cross section by 
$2.3$\%.  I regard this decrease as a $1\sigma$ uncertainty, which in
principle could either decrease or increase the cross section, 
since the 
magnitude of the forbidden correction is only an
estimate\cite{holstein86}. 

The uncertainty in the ${\rm^{51}Cr}$ Q-value causes an uncertainty of 
$\pm 0.05$\%, the ${\rm^{71}Ga}$ threshold 
$\pm 0.2$\%, and the ${\rm^{71}Ga}$ lifetime $\pm 0.3$\%.

Combining all of the uncertainties described above, I find an
effective $1\sigma$ uncertainty of $+3.6$\% ($-2.8$\%)
in the theoretical 
prediction of the cross section for absorption of ${\rm^{51}Cr}$ neutrinos
by ${\rm^{71}Ga}$.  
Three times the $1\sigma$ uncertainties quoted here are comparable to
the previously-estimated\cite{BU88} effective $3\sigma$ uncertainty of
$\pm 10$\%.
The excellent agreement 
with the measured
value GALLEX and SAGE 
values\cite{gallexcr1,gallexcr2,sagecr} is not significantly affected
by the 
recalculation described in this subsection.

The $3\sigma$ lower limit from excited state contributions that is 
adopted here,
namely, zero contribution, is absolutely reliable.  There is no way of
giving a similarly reliable {\it theoretical} 
upper limit  for the
contribution of excited states.  In fact, Hata and Haxton have argued 
that the only convincing upper limit is determined by the $^{51}$Cr 
measurements themselves.

\subsection{Constraint on Excited State Transitions}
\label{crcrossconstraint}

Anselmann {et al.}\cite{gallexcr1} and 
Hata and Haxton\cite{hata95} have shown how the measurements 
by GALLEX and SAGE of the 
capture rate by ${\rm^{51}Cr}$ neutrinos can be used as a constraint on 
the strength of the absorption transitions leading to the 
$5/2^-$ state at $175$ keV and the $3/2^-$ state at $500$ keV.
The measured capture rate, given in Eq.~(\ref{Rdef}),
Eq.~(\ref{BU88value}), and Eq.~(\ref{Rav}), can be written as the
ratio of the contributions to the ground state, the first excited
state at $175$ keV, and the excited state at $500$ keV, all divided
by the ground-state to the ground-state cross section.  The ratio of
the previously-standard cross section, Eq.~(\ref{BU88value}), to the 
best current value for the ground-state to ground-state transition,
$\sigma {\rm (^{51}Cr)_{\rm g.s.}} = 55.3 \times 10^{-46}~{\rm cm^2}$,
is $1.071$ . With a little algebra, one finds

\begin{equation}
\left[0.669\  {{\rm BGT(175~keV)}\over {\rm BGT}_{\rm g.s.}} + 0.220 
\ {{\rm BGT(500~keV)}\over {\rm BGT}_{\rm g.s.}}\right]
~=~ -0.004 \pm 0.075 ,
\label{crconstraint}
\end{equation}
where the ratio of the cross sections $\sigma({\rm 175~keV})/
\sigma_{\rm g.s.}$ would be $0.669$ if the BGT values for the two
transitions were equal [and $\sigma({\rm 175~keV})/
\sigma_{\rm g.s.}$ would be $0.220$ if the BGT values were equal].
Equation~(\ref{crconstraint}) implicitly assumes that the detection 
efficiency for the gallium experiments is close to unity; this 
assumption is based upon the independent tests for the detection
efficiency that are described in the original experimental 
papers\cite{gallexresults,sageresults}.

The coefficients in Eq.~(\ref{crconstraint}) are slightly different
from those given by Hata and Haxton, reflecting the slightly improved
data described in Sec.~\ref{input}.
In what follows, I shall use the prescription of the 
Particle Data Group\cite{PDG92} for estimating confidence limits,
namely, I shall assume that the errors shown in 
Eq.~(\ref{crconstraint}) are normally distributed with
a mean value at $-0.004$ but with a renormalized probability
distribution that is non-zero  only when the manifestly positive
quantity between the brackets in Eq.~(\ref{crconstraint}) is positive.

At the $3\sigma$ limit, Eq.~(\ref{crconstraint}) allows maximum values of 
${\rm BGT(175~ keV)}_{3 \sigma} = 12\times{\rm BGT(175~ keV)}_{p,n}$ and
${\rm BGT(500~ keV)}_{3 \sigma} = 6.8\times{\rm BGT(500~ keV)}_{p,n}$
where the values of ${\rm BGT}_{p,n}$ are described in
Sec.~\ref{excited}.
These maximum values of 12 and 7 times the indicated $(p,n)$ BGT
values are consistent with the constraints suggested by Hata and
Haxton.
These $3\sigma$ upper limits are probably  unrealistically conservative
since the available evidence  shows  
that $(p,n)$ reactions overestimate the
strength of two very weak beta-decay transitions (see
Sec.~\ref{cruncertainties} and Ref. \cite{hata95}). 
I have also implemented the $3\sigma$ limits in a conservative way:
I use the upper limit value for ${\rm BGT(500~ keV)}$ [with 
${\rm BGT(175~ keV)} = 0$] whenever the $500$ keV state 
is above threshold, since the ${\rm^{51}Cr}$ limit allows a somewhat larger
BGT value for a $500$ keV excited state than for a $175$ keV excited
state. 
I use the upper limit value for  ${\rm BGT(175~ keV)}$ below the 
threshold of the $500$ keV state.
This simplistic prescription causes a slight discontinuity in the 
upper error estimate near $500$ keV.

\section{Uncertainties Due to Excited State Transitions}
\label{exciteduncertainties}

I calculate
the  allowed $3\sigma$ lower limit for all neutrino
sources except the high energy ${\rm^8B}$ and $hep$ neutrinos by setting
equal to zero the
matrix elements for all excited state transitions. For the $3\sigma$ upper
limit to the weak transitions to the first two allowed excited states 
(Fig.~\ref{crdecayscheme}), 
I  follow Hata and Haxton\cite{hata95} and 
use the 
constraint provided by the ${\rm^{51}Cr}$ measurements that is given in 
Eq.~(\ref{crconstraint}). As discussed in Sec.~\ref{crcrossconstraint},
this ${\rm^{51}Cr}$ constraint
allows one of the transitions to be an order of magnitude
larger than indicated by the $(p,n)$ measurements; a difference this
large has not been
observed in any case in which accurate beta-decay measurements could
be compared with the results of $(p,n)$ experiments\cite{hata95}.
Indeed, for intrinsically 
weak transitions like the ones we are discussing, 
the $(p,n)$ experiments overestimate by an order of
magnitude the 
BGT values in the two  cases for which a comparison has been possible.

For higher energy neutrinos that can excite the many other GT
transitions, I determine as before  the $3\sigma$ lower limit by 
dividing by two the BGT values
determined by $(p,n)$ experiments and the $3\sigma$ upper limit by
multiplying the $(p,n)$ BGT values by a factor of two.
This is also an extreme range since for moderately strong transitions
a good correlation exists between the forward-angle $(p,n)$ cross
sections and the measured $\beta-$decay strength\cite{taddeucci87}.

The limits given later in this paper
represent maximal changes in that I assume that
all of the excited states are increased together, or decreased
together, to their extremal values.  I combine  linearly  
(rather than quadratically) the
uncertainties from different excited states.

\section{Solar Thermal Effects}
\label{thermaleffects}

The neutrino energies for laboratory weak interactions are shifted by
small amounts, $\Delta E$, due to the thermal 
energy of the particles that produce
the neutrino emission.  In general, one can write
\begin{equation}
q_{\odot} = q_{\rm lab} + \Delta E~,
\label{qsun}
\end{equation}
where $q_{\rm lab}$ and $q_\odot$ are, respectively, the neutrino
energies emitted in laboratory experiments and in the solar interior. 
The energy shifts are negligible 
for isolated nuclei, like ${\rm^8B}$, ${\rm^{13}N}$, ${\rm^{15}O}$, 
and ${\rm^{17}F}$,
that beta-decay in the sun.\cite{bahcall91}.
The physical reason for the smallness of the thermal effects in these
cases is that the initial decaying state contains only a single
nucleus,  unlike the situation in the $pp$ reaction in which two
fusing protons use their initial kinetic energetic to penetrate the
Coulomb barrier between them.  The random nature of the thermal
velocities causes all first order effects in the velocities of the
ions to vanish.  Therefore the change in shape is of order the 
temperature divided by the typical neutrino energy or keV over MeV.

However, when two or more particles are involved in neutrino
production, e.g., the  $pp$ reaction or ${\rm^7Be}$ electron capture, the
thermal shifts can produce small changes in the neutrino energies
and therefore neutrino absorption cross sections. These shifts due to
the addition of the stellar thermal energy to the laboratory decay energy 
are close to the level  of current experimental sensitivity.

The energy shifts, $\Delta E$, have previously been included in the energy
balance\cite{BU88}, at least in the nuclear energy generation code
that I have written and made publicly available\cite{website}.
However, the extra thermal energy has not been previously included in
the calculated absorption cross sections.   
I summarize  in this section the expected thermal energy shifts that
will be used in the following sections. The energy shifts have been
calculated in Refs.~\cite{BU88,bahcall91,bahcall94}.

The most important thermal contributions to the neutrino energy
spectrum are for the abundant, low-energy $pp$ neutrinos. An explicit
calculation for the average center of mass energy contributed by the
two fusing protons yields\cite{bahcall91}, Eq.~(52),

\begin{equation}
\Delta E (pp) = 3.6\times \langle d\phi_{pp}(T) 
\left(T/15 \times
10^6 K\right)^{2/3}\rangle\,\,{\rm keV},
\label{deppT}
\end{equation}
where $T$ is the ambient temperature.
The average indicated in Eq.~(\ref{deppT})
by the angular brackets is  taken over the temperature profile of
the sun weighted by the fraction of the $pp$ flux, 
$d\phi_{pp}(T)$,
that is produced at each temperature.
This average is insensitive to the details of the solar
model.  

I have calculated the average for four different
solar models: the 1992 Bahcall-Pinsonneault models\cite{BP92} 
with and without helium diffusion and the 1995 Bahcall-Pinsonneault
models\cite{BP95} with and without helium and heavy element diffusion.
The results for the four models can be summarized as follows:

\begin{equation}
\langle \, d\phi_{pp}(T) \left(T/15 \times
10^6 K\right)^{2/3} \rangle ~=~0.882 \pm 0.004~,
\label{weightedT}
\end{equation}
where the error shown is an indication of the systematic uncertainty. 
Thus the thermal energy shift for the $pp$ neutrinos is
\begin{equation}
\Delta E (pp) = 3.18 \pm 0.02\,\,{\rm keV} .
\label{depp}
\end{equation}
The variations in the energy shift from one solar model to another
that are 
indicated in Eq.~(\ref{weightedT}) are only about $20$ eV,
which is less than $0.01$\% of the total $pp$ energy.

For the much higher energy $hep$ neutrinos, the 
thermal energy shift is\cite{bahcall91}, Eq.~(53),
\begin{equation}
\Delta E (hep) = 5.4\times\langle   d\phi_{hep}(T) 
\left(T/15 \times 10^6\,\,K\right)^{2/3} \rangle \,\, {\rm keV}.
\label{hepshift}
\end{equation}
Using the same four solar models cited above in connection with the
evaluation of the weighted average $pp$ energy shift, I find for
the $hep$ thermal energy shift

\begin{equation}
\Delta E (hep) = 4.76 \pm 0.02\,\,{\rm keV} .
\label{dhep}
\end{equation}
The $20$ eV model dependence of the $hep$ energy shift is 
negligible compared to the $18.8$ MeV endpoint energy.

I have performed elsewhere\cite{bahcall94}
a detailed calculation of the shape of the 
energy profile of neutrinos produced by ${\rm^7Be}$ electron capture 
in the sun. The average energy for a neutrino emitted in the sun can
be expressed as the sum of the energy for a laboratory decay plus a
small correction due to the solar thermal energy. I find\cite{bahcall94}

\begin{subequations}
\begin{equation}
\langle \, q_{\odot} \left({\rm ^7Be}\right) \rangle ~=~
 861.8\,\,{\rm keV} +
1.28\,\,{\rm keV} = 863.1\,\,{\rm keV},\ \ 89.7\%
\label{q7bestrong}
\end{equation}
\begin{equation}
\langle \, q_{\odot} \left({\rm ^7Be}\right) \rangle  ~=~
 384.3\,\,{\rm keV} + 1.24\,\,{\rm keV}
= 385.5\,\,{\rm keV},\ \ 10.3\%
\label{q7beweak}
\end{equation}
\end{subequations}
When the energy shifts were evaluated using different solar
models, 
the spread in energy shifts was found to be 
less than $10$ eV. 

The energy shift for the $pep$ neutrinos has not yet been calculated
accurately, but this does not cause any significant uncertainty in the
predicted solar 
event rates since\cite{bahcall89}
 the calculated $pp$ flux is about $400$ times larger
than the $pep$ flux and the ${\rm^7Be}$ flux is about $40$ times larger.

\section{The \boldmath$\lowercase{pp}$ Neutrino Absorption Cross Section}
\label{ppsection}

In standard model calculations, 
the largest predicted contribution to the event rate of 
gallium solar neutrino experiments
is from the flux of $pp$ solar neutrinos\cite{bahcall89}.  
In this section, I reevaluate the neutrino
absorption cross section for $pp$ neutrinos including for the first
time the effect of the thermal energy of the fusing protons and making
use of the ${\rm^{51}Cr}$ constraint described in Sec.~\ref{crsection}.
I include overlap and exchange effects according to the prescription 
derived in Ref. \cite{bahcall63,bahcall78}. Exchange and 
overlap  effects reduce
the calculated $pp$ cross section by about $0.4$\% and by less than
$0.1$\%, respectively.

Appendix~A contains a numerical representation of the $pp$ neutrino
energy spectrum that should be sufficient for most particle physics
applications. In the calculations reported on here, I have used a more
detailed $pp$ spectrum, available elsewhere\cite{website}, that contains 
the relative intensity at 8000 different energies.

The q-value for the $pp$ reaction can be computed from the accurately
known and tabulated\cite{firestone96}
masses of the neutral atoms and is

\begin{equation}
q_{\rm lab} = 420.220 \pm 0.022 \, \, {\rm keV} .
\label{pplabq}
\end{equation}
Since hydrogen is ionized in the sun, the mass
difference for neutral atoms should be increased, for solar
calculations, by the binding energy of a K-electron,
$0.014$ keV.
Adding the thermal energy contributed by the fusing protons
[see Eq.~(\ref{depp}) or Ref.~\cite{bahcall91}] and the 
K-shell binding energy, 
one finds for the end point energy of 
the solar $pp$ energy spectrum

\begin{equation}
q_{\odot} = 423.41 \pm 0.03 \,\, {\rm keV} .
\label{ppsunq}
\end{equation}

The average energy loss 
from a star per emitted neutrino for an unmodified $pp$ energy
spectrum is an important datum for stellar evolution calculations and is
\begin{equation}
\langle q_{\odot}\rangle ~=~ 266.8 \,\, {\rm keV} .
\label{qppavsun}
\end{equation}

The best-estimate  ${\rm^{71}Ga}$  absorption 
cross section for $pp$ electron-type neutrinos with a standard 
model energy spectrum incident is

\begin{equation}
\sigma (pp) = 11.72\left[1.0  \pm 0.023\right] 
\times 10^{-46}\ \ {\rm cm^2}, \ 1\sigma. 
\label{ppcrosssection}
\end{equation}
Nearly all of the $pp$ cross section arises from ground-state to
ground-state transitions. Only $0.05$\% of the calculated cross
section given in Eq.~(\ref{ppcrosssection}) is estimated to arise from
transitions to the first-excited state of ${\rm^{71}Ge}$.

The value of the $pp$ cross section 
given in Eq.~(\ref{ppcrosssection}) is, 
as a result
of a number of cancelling changes, 
 only 0.7\% less than the previously calculated value\cite{BU88,bahcall89}. 
Adding the solar thermal energy to the
laboratory q-value increases the cross section by about
$1.6$\%.  The more precise evaluation of $\sigma_0$
(cf. Sec.~\ref{secsigmazero}), 
due primarily to improved calculations of the electron wavefunctions
and a more precise energy threshold, decreases $\sigma_0$ by about
\hbox{$2.2$\% .} The 
smaller threshold increases  the calculated  phase space,
and exchange effects reduce the capture cross section, both by
about 0.4\% .  Other corrections result in smaller changes.

The largest $1\sigma$
uncertainties in calculation of the $pp$ cross section
are caused by forbidden corrections ($\pm 2.3$\%), the matrix element 
from the ground-state of ${\rm^{71}Ga}$ to the first excited state of
${\rm^{71}Ge}$ ($+ 0.19$\%,  $- 0.02$\%), the lifetime of ${\rm^{71}Ge}$
($\pm 0.26$\%), and the ${\rm^{71}Ga}$-${\rm^{71}Ge}$ neutrino energy threshold
($\pm 0.1$\%).  I have taken the $\pm 1\sigma$ 
uncertainty due to forbidden
corrections to be equal to the decrease in the calculated cross
section when forbidden terms are set equal to zero.  The $3\sigma$
lower limit due to excited states was evaluated by setting 
equal to zero the matrix
element for the only excited-state transition 
that is energetically allowed.
The upper limit was determined by evaluating the maximum allowed
excited-state contribution that is consistent with the constraint
imposed by the ${\rm^{51}Cr}$ measurements [see Eq.~(\ref{crconstraint}].

\section{The $\mathbf{^8B}$, \boldmath$\lowercase{hep}$, 
$\mathbf{^{13}N}$, $\mathbf{^{15}O}$, 
and $\mathbf{^{17}F}$ Neutrinos}
\label{continuumcross}

The calculation of the absorption 
cross sections for neutrinos  from 
the beta-decaying sources, ${\rm^8B}$, ${\rm^{13}N}$, ${\rm^{15}O}$, 
and ${\rm^{17}F}$,
is simpler than for the $pp$ neutrinos because the shapes of the
neutrino energy spectra are changed by at most one part in $10^5$
by solar effects\cite{bahcall91}.

The effects of excited
state transitions are significant for all of the neutrino sources
considered in this section, again unlike the $pp$ reaction in which
the neutrino 
energy is too low for there to be  a significant probability for exciting
the final nucleus.

\subsection{$\mathbf{^8B}$ Neutrinos}
\label{b8continuum}

For the ${\rm^8B}$ neutrinos, I use the recently-determined best-estimate
neutrino spectrum\cite{bahcalllisi} and the improved gallium input
data(see Sec.~\ref{input}) 
to obtain a best-estimate absorption cross section of 

\begin{equation}
\sigma ({\rm ^8B}) = 2.40\left[1.0^{+ 0.32}_{- 0.15} \right] 
\times 10^{-42}\ \ {\rm cm^2}, \ 1\sigma. 
\label{b8crosssection}
\end{equation}
The value given in Eq.~(\ref{b8crosssection}) is essentially identical
to the 
cross section calculated in Ref.~\cite{bahcalllisi}.  However, the systematic 
uncertainties in this cross section
are large, because there
has not been any significant improvement in our understanding of the
relation between $(p,n)$ reactions and Gamow-Teller transition strengths.

The principal uncertainty in estimating the cross
section for ${\rm^8B}$ neutrinos is caused by the dominant contribution 
of excited states to the total cross section. For ${\rm^8B}$ neutrinos,
the ground-state to ground-state transition accounts for only $12$\%
of the best-estimate total cross section.  Since
transitions to many different 
excited states contribute to the ${\rm^8B}$ absorption cross
section, it is reasonable to assume that the stronger transitions
dominate. 
In estimating the uncertainties from
excited state transitions, I have taken advantage of the fact that
for relatively strong GT transitions the measured  
$(p,n)$ cross sections give BGT values that agree reasonably well with
BGT values determined directly from beta-decay\cite{hata95}. I have
therefore followed my usual practice\cite{bahcall89} and estimated 
the $3\sigma$
upper 
uncertainty for excited state transitions by doubling the contribution
estimated from $(p,n)$ reactions and the $3\sigma$ lower uncertainty by 
halving the calculated contribution from excited states.  The
uncertainty due to the shape of the ${\rm^8B}$ neutrino energy spectrum and
to forbidden corrections (see
Ref.~\cite{bahcalllisi}) are  much smaller, only $1.5$\% and 2.4\%,
respectively.

The neutrino energy spectrum from ${\rm ^8B}$ beta-decay does not have
a sharp cutoff because the predominantly-populated final state in
${\rm ^8B}$ is broad.  The average neutrino energy emitted is
\begin{equation}
\langle q_\odot\rangle\ =\ 6.735~{\rm MeV}\ \pm\ 0.036~{\rm MeV}~,
\label{avenergy}
\end{equation}
where the error estimate represents a $1 \sigma$ uncertainty, as
defined in Ref. \cite{bahcalllisi}, in the standard neutrino energy spectrum.

\subsection{CNO Neutrinos}
\label{cnocontinuum}

I recalculate the cross sections for CNO neutrinos in this subsection.
The changes from previous best-estimate
values \cite{BU88} are small in all cases.
The estimated uncertainties given here 
are larger than I previously estimated because I now use extreme criteria 
for determining the allowed range of contributions from excited state
transitions(see Sec.~\ref{exciteduncertainties} ).

In Appendix ~B, 
I present the calculated spectral energy distributions for the three
CNO neutrino sources.  These standard energy
distributions
are useful for many
particle physics applications, but I have not previously
published the CNO neutrino energy spectra.

The best-estimate absorption cross section for ${\rm^{13}N}$ neutrinos is 

\begin{equation}
\sigma ({\rm ^{13}N}) = 60.4\left[1.0^{+ 0.06}_{- 0.03} \right] 
\times 10^{-46}\ \ {\rm cm^2}, \ 1\sigma \ .
\label{n13crosssection}
\end{equation}
The cross section given
here corresponds to a spectrum with a maximum neutrino energy of 

\begin{equation}
q_{\odot} = 1.1982 \pm 0.0003  \,\, {\rm MeV} \ ,
\label{n13sunq}
\end{equation}
which is about $1$ keV smaller 
than I have used previously.  I have taken
account in the present calculation of the difference in binding energies
between initial and final neutral atomic states in the 
laboratory.  The average energy loss accompanying 
${\rm^{13}N}$ beta-decay is

\begin{equation}
\langle q_{\odot}\rangle ~=~ 706.3 \,\, {\rm keV} .
\label{qn13avsun}
\end{equation}

The upper-limit uncertainty is dominated  by our lack
of knowledge of the transition rates to excited states.  The $3\sigma$
upper limit is chosen so as to yield the maximum possible cross section 
for ${\rm^{13}N}$
neutrinos consistent with the constraint, Eq.~(\ref{crconstraint}),
from the ${\rm^{51}Cr}$ experiment. 
The uncertainty from forbidden corrections is added quadratically, but
is relatively small. 
The $3\sigma$ lower limit is determined
by setting  to zero 
the cross sections for all excited state transitions and by
multiplying by three the change in the cross section that results from
ignoring all forbidden corrections.  The effects of dropping excited
state transitions and of ignoring forbidden corrections have been
added quadratically.

The best-estimate absorption cross section for ${\rm^{15}O}$ neutrinos is 

\begin{equation}
\sigma ({\rm ^{15}O}) = 113.7\left[1.0^{+ 0.12}_{- 0.05} \right] 
\times 10^{-46}\ \ {\rm cm^2}, \ 1\sigma \ .
\label{o15crosssection}
\end{equation}
The maximum neutrino energy of the
${\rm^{15}O}$ neutrino energy spectrum is  
\begin{equation}
q_{\odot} = 1.7317 \pm 0.0005  \,\, {\rm MeV} \ ,
\label{o15sunq}
\end{equation}
after taking
account of the difference in binding energies
between initial and final neutral atomic states in the 
laboratory.  
The upper-limit and lower-limit uncertainties are determined
as described above for ${\rm^{13}N}$  neutrinos.
The average energy loss accompanying 
${\rm^{15}O}$ beta-decay is

\begin{equation}
\langle q_{\odot}\rangle ~=~ 996.4 \,\, {\rm keV} .
\label{qo15avsun}
\end{equation}

The cross section calculations 
for ${\rm^{17}F}$ neutrinos are  almost identical to those for
${\rm^{15}O}$ since the end-point energies are almost the same.  
I find

\begin{equation}
\sigma ({\rm ^{17}F}) = 113.9\left[1.0^{+ 0.12}_{- 0.05} \right] 
\times 10^{-46}\ \ {\rm cm^2}, \ 1\sigma \ .
\label{f17crosssection}
\end{equation}
The maximum neutrino energy is 
\begin{equation}
q_{\odot} = 1.7364 \pm 0.0003  \,\, {\rm MeV} \ ,
\label{f17sunq}
\end{equation}
and the  average neutrino energy emitted is

\begin{equation}
\langle q_{\odot}\rangle ~=~ 997.7 \,\, {\rm keV} .
\label{qf17avsun}
\end{equation}

\section{\boldmath$\lowercase{hep}$ Neutrinos}
\label{hepcontinuum}

The calculations for the rare $hep$ solar neutrinos are 
analogous to the calculations for the ${\rm^8B}$ neutrinos, which are
discussed in Sec.~\ref{b8continuum} except that the thermal energy
shift given in Eq.~(\ref{hepshift}) 
must be included for the $hep$ neutrinos.
I find

\begin{equation}
\sigma (hep) = 7.14\left[1.0^{+ 0.32}_{- 0.16} \right] 
\times 10^{-42}\ \ {\rm cm^2}, \ 1\sigma. 
\label{hepcrosssection}
\end{equation}
Only $7$\% of the best-estimate cross section is from 
ground-state to ground-state transitions.

The maximum neutrino energy is
\begin{equation}
q_\odot\ =\ 18.778~{\rm MeV}
\label{maxenergy}
\end{equation}
and the average neutrino energy is
\begin{equation}
\langle q_\odot\rangle\ =\ 9.628~{\rm MeV}~.
\label{avnuenergy}
\end{equation}

\section{The $\mathbf{^7B\lowercase{e}}$,
\boldmath$\lowercase{pep}$,
and $\mathbf{^{37}A\lowercase{r}}$  Absorption Cross Sections}
\label{be7pluspep}

In this section, I present the calculated cross sections for the solar
neutrino
lines from ${\rm^7Be}$ electron capture and from the $pep$ process, 
electron capture during
the $pp$ reaction.
I also calculate the cross section for absorption of neutrinos from a
${\rm^{37}Ar}$ laboratory radioactive source, which emits a neutrino with
a similar energy to the dominant neutrino line from ${\rm^7Be}$
electron capture in
the sun.

\subsection{The Neutrino Line from $\mathbf{^7B\lowercase{e}}$ Electron Capture}
\label{be7line}

The ${\rm^7Be}$ neutrinos are  the second most
significant contributor to the  calculated event rates in gallium
neutrino experiments, according to the predictions\cite{bahcall89} 
of the standard solar model and the 
standard electroweak theory.
The best-estimate cross section, weighted according to the branching
ratios indicated in Eq.~({\ref{q7bestrong}) and Eq.~({\ref{q7beweak}),
is
\begin{equation}
\sigma ({\rm ^{7}Be}) = 71.7\left[1.0^{+ 0.07}_{- 0.03} \right] 
\times 10^{-46}\ \ {\rm cm^2}, \ 1\sigma \ ,
\label{be7crosssection}
\end{equation}
which is about 2\% smaller than calculated previously\cite{BU88}.
  The inclusion of
the thermal energy of the interacting electron and ${\rm^7Be}$ ion
[cf. Eqs.~(\ref{q7bestrong}) and (\ref{q7beweak})]
increases the cross section by only $0.2$\% .  Excited state
transitions contribute approximately $6$\% of the total best-estimate
cross section given in Eq.~(\ref{be7crosssection}).

The uncertainties given  in Eq.~(\ref{be7crosssection}) represent, at
the $3\sigma$ limit, extreme values.  The $3\sigma$ lower limit ($-9\%$)
was obtained by setting equal to zero all excited state contributions
and by decreasing the best-estimate cross section 
by three times the calculated 
contribution from forbidden corrections. 
The uncertainties were added in quadratures.
The $3\sigma$ upper limit ($+ 21\%$) corresponds to maximizing
the  BGT value allowed, at $3\sigma$,
by the experimental constraint, Eq.~(\ref{crconstraint}), 
on the observed capture rate from chromium 
neutrinos. This maximization is equivalent to multiplying by
seven 
the $(p,n)$-estimate for the 
BGT value leading to the $500$ keV excited state in ${\rm^{71}Ge}$
(see Anselmann {\it et al.}\cite{gallexcr1} for a similar argument
using the initial results of the Gallex source experiment).
The smaller uncertainty due to forbidden corrections, 2.4\%
($1\sigma$), was combined
quadratically with the excited state uncertainty.

\subsection{The Neutrino Line from the \boldmath$\lowercase{pep}$ Electron Capture Reaction}
\label{pepline}

The flux of $pep$ neutrinos is about $400$ times less than the flux of
$pp$ neutrinos\cite{bahcall89}.  Hence, it is not necessary to know
accurately the cross section for 
$pep$ neutrino absorption  by ${\rm^{71}Ga}$,
which is fortunate since the uncertainties, dominated primarily by 
the unknown strengths of transitions to excited states, are 
relatively large.

The best-estimate cross section for the $pep$ reaction is
\begin{equation}
\sigma (pep) = 204\left[1.0^{+ 0.17}_{- 0.07} \right] 
\times 10^{-46}\ \ {\rm cm^2}, \ 1\sigma \ ,
\label{pepcrosssection}
\end{equation}
which is about 5\% smaller than calculated previously\cite{BU88}.
Much of this  change is due to uncertainty in where to
locate the $(p,n)$ transition strength that was
found experimentally\cite{krofcheck85,krofcheck87} to be 
somewhere between $1.0$ MeV and $1.50$ MeV excitation energy in 
${\rm^{71}Ge}$.  In my earlier calculations\cite{BU88,bahcall89}, I 
assumed that this transition strength was centered on the $1.095$ MeV 
excited state in ${\rm^{71}Ge}$. For the calculations in this paper, I have
assumed that strength is located at about $1.25$ MeV, at the midpoint
of the experimentally-allowed region and close to the exited state at 
$1.30$ MeV excitation energy in ${\rm^{71}Ge}$.  Moving the presumed
location of this transition
strength from an assumed excitation energy of $1.25$ MeV to an
excitation energy of $1.10$ MeV increases the calculated cross section
by $3.8$\%. 

About $18$\% of the best-estimate cross section arises
from excited state transitions.

The $3\sigma$ lower limit is determined
by setting equal  to zero 
the cross sections for all excited state transitions and by
multiplying by three the change in the cross section that results from
ignoring all forbidden corrections.
The $3\sigma$ upper limit was determined by: 1.) allowing the maximum
contribution from the first two excited states of ${\rm^{71}Ge}$ that is
consistent with the constraint [see Eq.~(\ref{crconstraint}) ] 
from the ${\rm^{71}Cr}$ experiment, 2.) locating
at $1.10$ MeV (the lowest possible energy) the 
excitation strength to  ${\rm^{71}Ge}$ excited states
observed to be 
between $1.0$ MeV and $1.50$ MeV excitation energy, and 3.) doubling the 
BGT value determined from the $(p,n)$ reactions for the other relevant 
excited states.  The different 
contributions to the uncertainty embodied in the 
$3\sigma$ upper limit were added
quadratically.

The amount of 
thermal energy that the combining electron and two protons
contribute, on the average, to the $pep$ neutrino energy has not been
calculated accurately.  Fortunately, this is unimportant for our
purposes. The neutrino endpoint energy neglecting thermal energies is 
known precisely and is $1.442232$ MeV, when account is taken of the 
extra $13.6$ eV binding energy that is included in the tabulations of
the neutral atoms masses. If we augment this nuclear mass difference
by the same amount of thermal energy, $5$ keV,  as for the $pp$
reaction, which is a plausible approximation, we obtain

\begin{equation}
q_{\odot}~\simeq~ 1.445 \,\, {\rm MeV} \ .
\label{pepsunq}
\end{equation}
The best-estimate cross section given in Eq.~(\ref{pepcrosssection})
was calculated using the approximate neutrino energy given in 
Eq.~(\ref{pepsunq}).  The calculated 
cross section is decreased by only $0.6$\%, of negligible importance
for solar neutrino experiments,
if the entire estimated  $5$ keV thermal energy is dropped.

\subsection{The Neutrino Line from $\mathbf{^{37}A\lowercase{r}}$ 
Electron Capture in the
Laboratory} 
\label{argonline}

Haxton\cite{haxton86} has suggested using a laboratory radioactive
source of ${\rm^{37}Ar}$ to test the efficiency of radiochemical solar
neutrino detectors. 
The neutrino energy of the ${\rm^{37}Ar}$ K-shell decay is $0.811$ MeV;
the L-shell energy is $0.813$ MeV.
Thus  neutrinos from  ${\rm^{37}Ar}$ decay 
in the laboratory have energies within 
several percent of the
energy [$863$ keV, see Eq.~(\ref{q7bestrong})] of the dominant 
${\rm^7Be}$ line.  As Haxton has emphasized, 
an experiment  carried out with an  intense ${\rm^{37}Ar}$
source would therefore 
provide a valuable additional test of the overall efficiency of 
gallium detectors in observing the important ${\rm^7Be}$ neutrinos.

The calculation of the  absorption cross section for ${\rm^{37}Ar}$
neutrinos is very similar to the calculation of the absorption cross
section for ${\rm^7Be}$ neutrinos described in Sec.~\ref{be7line}.  I find 

\begin{equation}
\sigma ({\rm ^{37}Ar}) = 70.0\left[1.0^{+ 0.07}_{- 0.03} \right] 
\times 10^{-46}\ \ {\rm cm^2}, \ 1\sigma \ ,
\label{ar37crosssection}
\end{equation}
which agrees well with the previously published value of $72 \times
10^{-46} \, {\rm cm^2}$\cite{bahcall89}. The
calculated 
total cross sections for ${\rm^{37}Ar}$ and ${\rm^7Be}$ neutrinos differ by only
2.4\% [cf. Eq.~(\ref{be7crosssection}) and Eq.~(\ref{ar37crosssection})].
As stressed by Haxton, if one varies the assumed energy of the dominant
excited state transition,
the $^{37}$Ar neutrino absorption 
cross section tracks the cross section  for $^7$Be neutrinos
remarkably well.  Considering three extreme cases, i.e., no excited state
transitions, the maximum allowed strength for the transition 
to the $175$ keV excited
state, and the maximum allowed strength to the $500$ keV excited
state, the total spread in the ratio of the $^7$Be to the $^{37}$Ar
neutrino absorption cross sections is only $0.6$\%.

The contributions of  the
two energetically---allowed transitions to 
excited states of ${\rm^{71}Ge}$ (see Fig.~\ref{gagefigure})
are proportional to the BGT
values for those excited state transitions. Thus one can write 
\begin{equation}
\sigma(^{37}{\rm Ar}) ~=~\left[66.2 ~+~ 
46.0\  {{\rm BGT(175~keV)}\over {\rm BGT}_{\rm g.s.}} ~+~
17.4\ {{\rm BGT(500~keV)}\over {\rm BGT}_{\rm g.s.}}\right] \times10^{-46}\,
{\rm cm^{-2} }.
\label{arcontributions}
\end{equation}
The best-estimate cross section given in Eq.~(\ref{ar37crosssection}) 
was obtained by
using in Eq.~(\ref{arcontributions}) 
the BGT values indicated by $(p,n)$ reactions (see Sec.~\ref{input}
and Table~\ref{BGT}).
 Excited state transitions
contribute $5$\% of the best-estimate cross section for ${\rm^{37}Ar}$
neutrinos, very similar to the  $6$\% contributed by excited states 
to the best-estimate ${\rm^7Be}$ neutrino absorption cross section. 
The uncertainties in the calculated cross section that are shown in 
Eq.~(\ref{ar37crosssection}) were calculated in the same way as for
the ${\rm^7Be}$ line (see Sec.~\ref{be7line}).

\section{Absorption Cross Sections at Specific Energies}
\label{crossspecific}

Neutrino absorption cross sections at specific energies 
are required in order
to calculate the capture
rates predicted by different  scenarios with new physics (e.g., 
neutrino oscillations with a variety of mixing 
parameters) in which the energy
spectrum of solar neutrinos is changed
from the standard neutrino energy spectrum.
In this section, I provide the required cross sections as a function of
energy  and, for the first
time, also  present the uncertainties in the cross sections as a
function of energy.

Table~\ref{bestsigmacue} gives the best-estimate neutrino cross
sections at a set of strategically chosen energies.
The cross sections were evaluated 
according to the precepts described in the
previous sections.
Using a cubic spline fit\cite{press92}
 to the cross sections as a function
of energy that are given in Table~\ref{bestsigmacue}, I have verified that
the numbers given in the table are sufficient to 
reproduce to an accuracy of 1\% or better 
the best-estimate cross sections calculated in the previous sections
for the standard neutrino energy
spectra.

The uncertainties in the neutrino cross sections depend upon 
neutrino energy
since the number of accessible excited states increases with energy
and the forbidden corrections also increase with energy.
The energy dependence of the cross section uncertainties 
has not, so far as I know,  been taken into account in the previously 
published
comparisons of the calculated and observed rates in neutrino
experiments. 
In order to provide the data with which to include the
energy-dependent cross section uncertainties in future analyses,
I have combined quadratically the uncertainties from different
sources, as indicated in the discussion in the previous sections, and
have calculated $3\sigma$ upper and lower limit cross sections at the
same energies at which cross sections are listed in
Table~\ref{bestsigmacue}.

The $3\sigma$ limit cross sections 
are given in Table~\ref{minsigmacue} and
Table~\ref{maxsigmacue}. 
I have also verified that a cubic spline fit to 
the cross sections given 
in these tables may be used to
evaluate the cross section uncertainties in the rates
predicted by physics scenarios with non-standard  neutrino energy
spectra.

Figure~\ref{sigmaque} shows the calculated cross sections, and the
estimated $3\sigma$ uncertainties, as a function of neutrino energy.
For energies above $25$ MeV, the uncertainties become so large as to
make the calculated cross sections not very useful.

\section{Predicted Solar Neutrino Event Rates}
\label{eventrates}

The event rates measured by the GALLEX\cite{gallexresults} and 
SAGE\cite{sageresults} solar neutrino
experiments are significantly less than the 
standard solar model predictions if nothing happens to the neutrinos
after they are produced in the center of the sun.  How can we assess
the significance of this deficit?

Over the years, I have given a formal quantitative measure of the
reliability of the theoretical predictions by determining errors in  the
calculations based upon the recognized uncertainties in the input
data.  This work has been published largely in the Reviews of Modern
Physics and the investigations prior to 1989 are summarized in
Neutrino Astrophysics\cite{bahcall89} (for recent improvements see
\cite{BP92} and \cite{BP95}).

In this section, I  provide two different ways of assessing the robustness
of the theoretical predictions. In Sec.~\ref{standardrates}, I 
review all of the published standard solar model calculations since 1963 in
which my colleagues and I have been involved. The variation over time 
of the standard model neutrino fluxes 
provides an  intuitive feeling for the reliability of the
theoretical calculations. In Sec.~\ref{minimalrates}, I present
the results of a series of new solar model calculations in which
different nuclear reaction rates are set equal to zero in order to
minimize artificially the calculated total event rate for gallium
neutrino experiments. These solar model {\it gedanken} experiments
provide a different 
 indication of how difficult it is to lower significantly
the predicted solar model event rates.

\subsection{Standard Solar Model Predictions}
\label{standardrates}

Figure~\ref{gahistory} shows the event rates computed for all the
neutrino fluxes predicted by the then best standard
solar models which I and my collaborators have published
since the first such model appeared in 1963~\cite{fibs}.
To isolate the effects of solar models, the rates shown in
Fig.~\ref{gahistory} were computed in all cases with the  absorption cross
sections determined 
in the present paper for standard solar neutrino energy spectra.  
The uncertainties indicated in Fig.~\ref{gahistory} are the
$1\sigma$ errors due just to the cross section uncertainties estimated  in
the present paper.  I have assumed that the uncertainties from
different excited state transitions add linearly and coherently, 
i.e., the cross sections for the individual neutrino sources are 
simultaneously increased (or decreased) to their  maximum 
(minimum) allowed values. 

The predicted neutrino fluxes have been remarkably constant in time over
the last three decades.  
Prior to this time, 
in the first several years of solar neutrino
studies that are represented by the  earliest points in
Fig.~\ref{gahistory}, 
the cross sections for the low energy nuclear
physics reactions were not well known
and the reaction rates
calculated with the then-current nuclear cross sections led to large
values for the higher energy, more easily detectable neutrinos.
In 1964, when the chlorine solar neutrino experiment was
proposed\cite{davis64,bahcall64}, the rate of the ${\rm^3He}$-${\rm^3He}$
reaction was estimated\cite{good54,parker64} to be $5$ times slower
than the current best-estimate and the uncertainty in the low energy
cross section was
estimated\cite{parker64}  
to be ``as much as a factor of $5$ or $10$.''  Since the 
${\rm^3He}$-${\rm^3He}$ reaction competes with the 
${\rm^3He}$-${\rm^4He}$ reaction---which leads to high energy neutrinos---the
calculated fluxes for the higher energy neutrinos were overestimated in
the earliest days of solar neutrino research.
The most significant uncertainties, in the rates of the 
${\rm^3He}$-${\rm^3He}$,
the ${\rm^3He}$-${\rm^4He}$, and the ${\rm^7Be}$-$p$ reactions, were much reduced after
just a few years of intensive experimental research in the middle and late
1960s\cite{bahcalldavis82}.

The event rates for gallium appear even more robust when account is taken
of the fact that prior to 1992 the standard  solar models did not
include  the
effects of diffusion.  Using cross sections calculated in this paper
and neutrino fluxes predicted by 
the 1995 Bahcall-Pinsonneault\cite{BP95}
standard model (which
includes helium and heavy element diffusion and the 1995
best-estimates for the nuclear reaction rates), as well as recent 
improvements in radiative opacity and equation of
state\cite{bahcall97}, the
calculated  event rate is 

\begin{equation}
{\rm Standard~Solar~Model~Capture~Rate~With~Diffusion ~=~ 135 ~\, SNU},
\label{bestgarate}
\end{equation}
$1$ SNU less than calculated with the previously-used cross
sections\cite{BP95}. 
Omitting diffusion, but otherwise using all the same code and data to
construct a standard solar model\cite{BP95}, the calculated rate with
the gallium cross sections given in this paper  is

\begin{equation}
{\rm Standard~Solar~Model~Capture~Rate~Without~Diffusion ~=~ 124 ~\, SNU}.
\label{nodiffgarate}
\end{equation}
Comparing the results given in Eq.~(\ref{bestgarate}) and
Eq.~(\ref{nodiffgarate}), we see that 
the effect of including diffusion is to increase by about
\hbox{$11$~SNU} 
the standard
solar model prediction for the gallium solar neutrino event rate.

Helioseismological measurements show that element diffusion is
occurring in the sun, confirming theoretical expectations.  The present-day
surface abundance of helium calculated from  solar models is
 in excellent agreement with the 
helioseismologically determined value only if diffusion is
included\cite{BP95};  the comparison of the computed and observed
depth of the convective zone also requires that diffusion be included
in the solar models\cite{BP92,BP95}.  More recently, it has been
shown\cite{bahcall97} that the sound velocities of the sun determined
by helioseismological measurements from $0.05 R_{\odot} $ to $0.95
R_{\odot}$ agree to within $0.1$\% rms
with the sound velocities calculated from a standard
solar model provided that diffusion is included in the model
calculations.  The mean squared discrepancy for a model without
diffusion is 22 times larger than for the standard model with
diffusion, indicating that models without diffusion are inconsistent
with helioseismological measurements\cite{bahcall97}.

If the values 
 prior to 1992 in Fig.~\ref{gahistory} are increased by $11$~SNU 
to correct for the omission of diffusion, then the corrected values
 since 1968 through 1997 all lie in the range $120$ SNU to  $141$~SNU,
i.e.,

\begin{equation}
{\rm Total~Historical~Range~Corrected~for~Diffusion}~=~ \hbox{120~SNU--141~SNU},
~~\hbox{1968--1997}.
\label{averagegarate}
\end{equation}

The observed event rate in the GALLEX detector is\cite{gallexresults}

\begin{equation}
{\rm GALLEX~Observation} ~=~ 70~ \pm 8 {\rm ~SNU},
\label{observedgallex}
\end{equation}
and the rate observed by the SAGE detector is\cite{sageresults}
\begin{equation}
{\rm SAGE~Observation} ~=~ 72~ \pm 13 {\rm ~SNU}.
\label{observedsage}
\end{equation}
The weighted average observed rate is $70.5 \pm 7$ SNU.

The difference between the predicted rate, Eq.~(\ref{bestgarate}), and
the observed rates, Eq.~(\ref{observedgallex}) and
Eq.~(\ref{observedsage}), is the essence of the contemporary 
gallium solar neutrino
problem.  Moreover, the GALLEX observation is, by itself, more than
$5\sigma$ below all the standard solar model results shown in
Fig.~\ref{gahistory} since 1968, if the values prior to 1992 are
corrected for the effects of diffusion. 

\subsection{Models With Selected Nuclear Reaction
 Rates Set Equal to Zero}
\label{minimalrates}

The neutrino absorption cross sections are a monotonically increasing
function of energy (cf. Table~\ref{bestsigmacue}). Therefore, the
minimum conceivable event rate is achieved if one artificially sets
equal to zero the nuclear reactions that produce higher energy
neutrinos, like the ${\rm^7Be}$, ${\rm^8B}$, and CNO neutrinos.  The first such
calculation was performed by Bahcall, Cleveland, and
Davis\cite{bahcall85}, who minimized the rate subject only to the
condition that the nuclear energy released by fusion in the solar
interior equal the present-day solar luminosity. Allowing only $pp$
and $pep$ neutrinos and using the previous best-estimates for gallium
neutrino absorption cross sections, 
these authors obtained a minimum allowed rate
---for standard neutrino physics---of $80$ SNU.

In Table~\ref{minrates}, I summarize the results of a series of solar
model calculations that were made by setting equal to zero selected
nuclear reactions. The models were constructed in the same way as the
best Bahcall-Pinsonneault solar models\cite{BP95}, except that specific
nuclear reactions were artificially set equal to zero in the nuclear
energy generation subroutine.  

The most dramatic decrease in the predicted gallium
event rate is achieved by setting to zero the rate of the well-known 
${\rm^3He(^4He},\gamma){\rm^7Be}$ reaction, which 
leads in the standard solar model 
to the neutrinos from ${\rm^7Be}$ electron and proton capture, the so-called
${\rm^8B}$ and ${\rm^7Be}$ neutrinos. With this reaction equal to zero, the only
way in the $pp$ chain of completing the fusion of protons into alpha
particles is by the low energy $pp$ reaction, with an occasional 
($\sim 0.2$\% by neutrino flux) $pep$ reaction.  

The calculated event rate is

\begin{equation}
{{\rm No~^3He(^4He},\gamma){\rm^7Be~Reactions}}~=~88.1^{+3.2}_{-2.4}~{\rm SNU},
\label{nohe3he4}
\end{equation}
with $1\sigma$ errors on the neutrino
absorption cross sections.
The corresponding rate in the chlorine solar neutrino
experiment is $0.73 \pm 0.01$~SNU, which is more than $10\sigma$ 
less than  the
observed rate\cite{chlorine} of $2.54 \pm 0.20$~SNU.
The rate,  $0.0$~SNU, predicted 
for the Kamiokande solar neutrino experiment
is $8\sigma$ less than the observed rate in the Kamiokande
experiment\cite{kamioka}. 
Nevertheless, the gallium event rate of $88$~SNU 
calculated in this concocted (clearly incorrect) model is
about 2.5 standard deviations larger than the combined observed 
rate of $70.5$  SNU in the GALLEX and SAGE experiments.

The primary reason that the rate given here is larger than the value
calculated by Bahcall, Cleveland, and Davis\cite{bahcall85} is that
some of the solar luminosity  and neutrinos are coming from the CNO fusion
reactions. The calculated event rate in the gallium experiments can be
reduced somewhat further 
if one sets equal to zero simultaneously the rates of
both the 
${\rm^3He(^4He},\gamma){\rm^7Be}$ reaction and the 
${\rm^{12}C}(p,\gamma){\rm^{13}N}$ 
reactions. In this case, the CNO neutrinos (from ${\rm^{13}N}$, 
${\rm^{15}O}$,
and ${\rm^{17}F}$ decays ) 
are all greatly reduced in flux and the 
${\rm^7Be}$ and ${\rm^8B}$ neutrinos are completely absent.  
Table~\ref{minrates} shows
the the calculated rate for this case is \hbox{$79.7^{+2.4}_{-2.0}$ SNU}.  

The minimum rate is achieved by simultaneous setting equal to zero
the reaction rates for the ${\rm^3He(^4He},\gamma){\rm^7Be}$ reaction and {\it
all} the CNO reactions.\footnote{In this case, one requires
zero cross sections for
${\rm^3He(^4He},\gamma){\rm^7Be}$, ${\rm^{12}C}(p,\gamma){\rm^{13}N}$,   
${\rm^{13}C}(p,\gamma){\rm^{14}N}$, ${\rm^{14}N}(p,\gamma){\rm^{15}O}$,
and
${\rm^{15}N}(p,\gamma){\rm^{16}O}$.}
In this case, I find

\begin{equation}
{\rm Minimum~Rate}~=~79.5^{+2.3}_{-2.0}~{\rm SNU},
\label{minrateequation}
\end{equation}
where the uncertainties are again $1\sigma$ errors. 
This extreme hypothesis also predicts $0.0$~SNU for the Kamiokande
experiment and $0.3$~SNU
for the ${\rm^{37}Cl}$ experiment, the latter
value is $11$ standard deviations less than
the observed capture rate in the chlorine detector\cite{chlorine}.  

The solar neutrino fluxes that produce  the minimum  neutrino
capture rate in gallium detectors 
are \hbox{$\phi(pp)~=~6.50\times10^{10}\, {\rm cm^{-2}s^{-1}}$},
\hbox{$\phi(pep)~=~1.61\times10^{8}\, {\rm cm^{-2}s^{-1}}$},
and \hbox{$\phi(hep)~=~1.4\times10^{3}\, {\rm cm^{-2}s^{-1}}$}.

\section{Summary and Discussion}
\label{summary}

I first summarize 
 the calculations of  neutrino absorption cross sections
and then discuss the 
event rates predicted by standard solar models and by extremely
 non-standard solar models.   In the last subsection, I suggest an
 answer to the question: What will GNO show?

I present in Appendix~\ref{appendixa} 
the standard $pp$ neutrino energy spectrum that was calculated with
the 
inclusion of  the thermal energy of  fusing ions and also present in 
Appendix~\ref{appendixb} the
standard CNO neutrino energy spectra.

\subsection{Cross Sections}
\label{crossdiscussion}

\subsubsection{Best Estimates}

Table~\ref{crosssummary} summarizes the best-estimates,
and the $1\sigma$ uncertainties, of the  neutrino
absorption cross sections that were calculated in the preceding sections 
for ${\rm^{71}Ga}$ targets. All of the cross sections given in
Table~\ref{crosssummary} were evaluated  assuming standard $\nu_e$
energy spectra and 
the input data for the ${\rm^{71}Ga}$-${\rm^{71}Ge}$ system
that are summarized in Sec.~\ref{input}.
The $f$-value for ${\rm^{71}Ge}$ electron capture used here makes use of
new Dirac-Fock self consistent field calculations of the electron wave
functions that include finite nuclear size, the Breit interaction,
and the most important QED corrections\cite{grant,johnson,chen}.  
In addition, 
I have  taken account 
of atomic overlap and exchange effects 
(see Sec.~\ref{smalltheory}), effects for which I have previously only
estimated upper limits.

I have also included here for the first time 
the thermal energy contribution of the fusing
particles to the neutrino absorption cross sections. 
Section~\ref{thermaleffects} presents the results of 
calculations of 
the thermal energy from  neutrino-producing reactions that occur in
the solar interior.  This  thermal energy increases 
by $1.6$\% the calculated
absorption  cross section for $pp$ neutrinos, 
but is unimportant for all other 
cases considered in this paper.  

The cross section for $\nu_e$ absorption by ${\rm^{51}Cr}$ calculated here
agrees to better than $1\sigma$
with the independent measurements by the
GALLEX\cite{gallexcr1,gallexcr2} and SAGE\cite{sagecr}  
experiments.  However, the 
measured rates in the GALLEX and SAGE solar neutrino experiments
differ by more than $5\sigma$ from all the 
standard solar model predictions 
of my colleagues and myself since 1968 provided the 
values published prior to 1992 
are corrected, as required by helioseismology, for 
the effects of diffusion(see Sec.~\ref{standardrates}).

\subsubsection{Particle Physics Applications}

Many particle physics explanations of the observed solar neutrino
event rates imply that the energy spectrum of $\nu_e$ solar neutrinos
is altered by new physics.  
In order to calculate the rates expected
from the variety of proposed 
non-standard neutrino energy spectra, one must have available neutrino
cross sections, and their uncertainties, 
 as a function of neutrino energy. I have not previously published 
uncertainties in the cross sections calculations as a function of
energy.  Therefore, in the many papers in which empirical analyses of
solar neutrinos were made using non-standard neutrino energy spectra,
the theoretical errors in the cross section calculations were, of
necessity,  either
ignored or (incorrectly) set equal to the published uncertainties for
cross sections with standard $\nu_e$ energy spectra.

Table~\ref{bestsigmacue} presents the required neutrino cross sections
at a set of neutrino energies that were chosen to permit,
with the aid of a cubic spline fit,
accurate
calculations for any specified neutrino energy spectrum.
I have also calculated $3\sigma$ different minimum and maximum
absorption cross sections, which are presented in
Table~\ref{minsigmacue} and Table~\ref{maxsigmacue}.

Figure~\ref{sigmaque}
displays the cross sections, and their uncertainties, as
a function of neutrino energy.
Using the best-estimates and $3\sigma$ different cross sections given
in the tables, one can calculate the uncertainties in predicted
neutrino event rates for an arbitrarily changed solar neutrino energy
spectrum.

\subsubsection{Uncertainties}

Since gallium solar neutrino experiments test fundamental aspects of
physics and astrophysics, I have adopted extremely conservative
criteria for the estimated uncertainties. In most cases, the largest
uncertainties arise from the poorly known strengths of transitions to
excited states in ${\rm^{71}Ge}$.   

For low and moderate energy neutrino sources, 
the $pp$, $pep$,  ${\rm^7Be}$, ${\rm^{13}N}$,
${\rm^{15}O}$, and ${\rm^{17}F}$ neutrinos,
I have set equal to zero the matrix elements
for all excited state transitions in order to determine the $3\sigma$
minimum cross sections.
I have calculated the $3\sigma$ upper limit for
the important 
transitions to the $175$ keV and the $500$ keV states in ${\rm^{71}Ge}$
(see Fig.~\ref{crdecayscheme}),
by using the constraint from the
GALLEX\cite{gallexcr1,gallexcr2,cribier88} and SAGE\cite{sagecr}  
measurements of
the ${\rm^{51}Cr}$ absorption rate [see Eq.~(\ref{crconstraint})].  
This prescription results 
in $3\sigma$ upper limit cross sections that are,
respectively, 7 and 12 times the values inferred using the measured
$(p,n)$ cross sections (see discussion in 
Sec.~\ref{crcrossconstraint}), which are  probably 
unrealistically large uncertainties 
since  the BGT values inferred from $(p,n)$
measurements have never been observed to exceed by large factors 
the true BGT values
determined by beta-decay experiments.

I have also taken a skeptical attitude toward the calculated 
values of the forbidden
corrections, and have adopted   $1\sigma$ uncertainties 
from forbidden effects equal to
the best-estimate values for the forbidden corrections.
For the low and moderate energy neutrino sources, the forbidden
corrections are always between $2$\% and $2.5$\%.
The $1\sigma$ uncertainties estimated in this way are significant
for the $pp$, ${\rm^7Be}$, ${\rm^{13}N}$, ${\rm^{37}Ar}$, 
and ${\rm^{51}Cr}$ 
neutrino cross sections
(cf. Table~\ref{crosssummary} ) but, with the exception of the $pp$
cross section, only for the lower limit value.

The largest uncertainties in the ${\rm^{51}Cr}$ calculation are from 
excited state transitions ($+2.8$\% and $-1.6$\%, $1\sigma$) and 
forbidden corrections ($\pm 2.3$\%, $1\sigma$).
I have also calculated the cross section for absorption of ${\rm^{37}Ar}$
neutrinos, since, as  Haxton\cite{haxton86} has discussed, the close
similarity between the ${\rm^{37}Ar}$ and ${\rm^7Be}$ neutrino energies makes
argon a theoretically attractive possible calibrator for the detection
efficiency for ${\rm^7Be}$ neutrinos.

\subsubsection{Correlations Among the Uncertainties}
\label{correlations}

How are the uncertainties correlated 
 between cross sections calculated for  different energies?
Some of the sources of uncertainties are fully correlated, e.g, the 
the characteristic $\sigma_0$ defined by Eq.~(\ref{defsigmazero}) 
and Eq.~(\ref{sigmazero})
is a common scale factor for all the cross sections.  On the other hand,
some sources of uncertainties are uncorrelated; uncertainties in
matrix elements to highly excited states in $^{71}$Ge affect the cross
sections for 
higher energy neutrinos but do not influence the cross sections
for lower energy neutrinos. 

I recommend the most conservative procedure: assume all errors are
fully correlated and add the uncertainties linearly not quadratically.
This is the procedure that I have followed in calculating 
Table~\ref{minsigmacue} and Table~\ref{maxsigmacue}.
For standard model predictions, i.e., standard solar
models and non-oscillating neutrinos, adding the uncertainties
linearly and quadratically will give approximately the same answer
because the uncertainties are dominated by the higher energy neutrinos
from $^8$B.  However, for non-standard neutrino scenarios, such as the MSW
effect or vacuum neutrino oscillations, the two procedures may give
significantly different results. To test for the sensitivity of the
error estimate to the prescription adopted, one can combine the errors
quadratically and also linearly and compare the difference error
estimates. In certain cases, it may be reasonable to break up the
calculations into different energy groups, e.g.,  below or
above $2$ MeV, and assume that the uncertainties are correlated within
each group  but not between groups.

\subsection{Predicted Event Rates}
\label{eventdiscussion}

Figure~\ref{gahistory} shows the event rates calculated using 
 all of the
standard solar model neutrino
fluxes that my colleagues and I have published since the first solar
model calculation of neutrino fluxes in 1963.  
In order to isolate the effect of the solar model predictions, 
I have used the 
absorption cross sections derived in this paper 
for all the points plotted.
The event rates have been  remarkably constant, especially since 1968 by
which time  the largest initial
uncertainties in determining the nuclear fusion 
cross sections were greatly reduced\cite{bahcalldavis82}.

In the $35$ years that we have been calculating neutrino fluxes from
standard solar models, many improvements have been made in the
in the input data for the solar models (in e.g., the 
nuclear reaction rates, opacities,
equation of state, and heavy element abundances) and in
the sophistication and precision of the stellar evolution codes (e.g.,
the inclusion of diffusion).
Throughout this whole period,
the historically lowest rate corresponds, without diffusion,  
to $109$ SNU (see the
lowest point in  
Fig.~\ref{gahistory}, which occurs in 1969), which is more than $5\sigma$
larger than the combined experimental result from the GALLEX and SAGE
experiments.

Detailed calculations and  helioseismological
measurements both
show\cite{BP92,BP95,bahcall97} 
 that we must correct for the effects of 
diffusion the neutrino fluxes calculated
prior to 1992
(cf. discussion in 
Sec.~\ref{standardrates}), 
All of the standard model fluxes that my colleagues and I have
calculated in the  $30$ years since $1968$
lie in the range $120$ SNU to  $141$ SNU, if the effects of diffusion are
included.  

The disagreement is robust between the predictions of standard solar
models--supplemented by the assumption that nothing happens to the
neutrinos after they are produced--and the results of gallium solar
neutrino experiments.

How much could one conceivably reduce the calculated  event rate in
gallium neutrino experiments assuming standard neutrino physics, but
non-standard (or impossible) nuclear physics? 
The most efficient way to reduce artificially the calculated  counting rate is 
 by setting equal to zero the nuclear reaction
rates that lead to higher energy neutrinos in the solar model
computations.
This arbitrary procedure is  physically impossible (the relevant nuclear 
fusion rates are
measured in the laboratory to be comparable to rates that are not set
equal to zero), but illustrates the
extreme difficulty in reducing 
 the calculated event rate to a rate close to what is measured
in gallium solar neutrino experiments. 

If we artificially eliminate all nuclear reactions that lead to 
${\rm^7Be}$ and ${\rm^8B}$ neutrinos (i.e., assume that the  
cross section measured
in laboratory experiments for  the
${\rm^3He}(\alpha,\gamma){\rm^4He}$ reaction is completely wrong and the reaction is
forbidden for some unknown reason), then the rate calculated from
standard solar models is $88.1^{+3.2}_{-2.4}$ SNU.  This rate is 
about $2.5~\sigma$ larger than 
the measured rate in the GALLEX and SAGE experiments. Moreover, the 
same solar model predicts an event rate of $0.73 \pm 0.01$ SNU for the
chlorine experiment, which is more than $8\sigma$ less than is
observed.

One can consider solar models in which obviously incorrect assumptions
about nuclear reactions are made for a number of different fusion
reactions. If one assumes that not only does the
${\rm^3He}(\alpha,\gamma){\rm^4He}$ reaction not occur, but also all four of the
$(p,\gamma)$ reactions in the CNO cycle (see footnote~2) 
do not occur, then one can calculate a
solar model for which the capture rate is $79.5^{+2.3}_{-2.0}$ SNU.
This capture rate corresponds to the minimum rate that is possible,
ignoring all of the physics of solar models and making false
assumptions about nuclear reactions, if one requires that the sun is
currently producing thermal 
 energy from nuclear fusion at the rate at which it is
radiating energy via photons escaping from its surface.
Even this most extreme model predicts an event rate that exceeds the
current best-estimate rate, about $70.5 \pm 7$ SNU,  observed by 
GALLEX\cite{gallexresults} and SAGE\cite{sageresults}.
The chlorine rate predicted by this most extreme model is $0.3$ SNU,
which is inconsistent with the observed rate of $2.54 \pm 0.20$
SNU\cite{chlorine}.  

\subsection{What Will GNO Show?}
\label{gnoshow}

Gallium solar neutrino experiments are the only established way of 
detecting the great majority of solar neutrinos, the low energy
neutrinos from the fundamental $pp$ reaction. Therefore, astronomical
inferences and particle physics 
applications of solar neutrino studies rely
strongly on the measured rates in gallium experiments. 
These inferences and applications will become  more
stringent as GNO reduces the statistical and the systematic
uncertainties in the measured gallium rate.

The most exciting result that GNO might obtain is, in my opinion, to
find a capture rate that is more than $3\sigma$ smaller than the 
minimum rate, $79.5^{+2.3}_{-2.0}$ SNU, 
 calculated in Sec. \ref{minimalrates}.
This minimal rate is obtained  by setting equal to zero
five well-measured (and appreciable) nuclear reaction rates, and
ignoring everything we know about the sun except its total luminosity.
If a number significantly  less than $80$ SNU were obtained, 
I believe that GNO by itself would establish that we require 
non-standard neutrino physics 
in order to explain solar neutrino experiments. 
In fact, I think the same conclusion would be drawn if 
GNO ruled out by more than $3\sigma$ 
the unrealistically low rate of $88.1^{+3.2}_{-2.4}$ SNU
obtained by arbitrarily excluding the nuclear reactions 
${\rm^3He}(\alpha, \gamma){\rm^7Be}$ that lead to 
${\rm^7Be}$ and ${\rm^8B}$ neutrinos.  

Will GNO find, after improvements in the statistics and in the
systematic errors, that the best-estimate capture rate is
significantly less than $88$ SNU, or even less than $80$ SNU?  
No one really knows, which is one of the reasons why
the experiment is so important.

\section*{Acknowledgments}
I am grateful to T. Bowles, 
B. Cleveland, A. Dar, S. Elliott, R. B. Firestone, 
 N. Fortson, W. Hampel,
W. Haxton, N. Hata, P. Krastev, P. Kumar,
K. Lande, W. Marciano,  J. Rappaport, and L. Wolfenstein
 for valuable comments or 
discussions. I am indebted to I. P. Grant, W. R. Johnson, and M. Chen
for performing Dirac-Fock self-consistent field calculations of the
electron wave functions in ${\rm ^{71}Ge}$. 
This research is supported by  NSF grant \#PHY95-13835

\appendix
\section{}
\label{appendixa}
Table~\ref{ppnumerical} gives the normalized unmodified energy 
spectrum, $P(q)$, 
for the $pp$ neutrinos.  The end point energy 
includes the average thermal energy of the fusing protons.
\bigskip
\medskip

\section{}
\label{appendixb}
Table~\ref{13Nnumerical} gives the normalized unmodified 
energy spectrum, $P(q)$, 
for the ${\rm ^{13}N}$ neutrinos.

Table~\ref{15Onumerical} gives the normalized unmodified 
energy spectrum, $P(q)$, 
for the ${\rm ^{15}O}$ neutrinos.

Table~\ref{17Fnumerical} gives the normalized unmodified 
energy spectrum, $P(q)$, 
for the ${\rm ^{17}F}$ neutrinos.

\begin{table}[htb]
\centering
\tighten
\begin{minipage}[h]{2.3in}
\caption[]{Gamow-Teller strength functions in ${\rm^{71}Ge}$ as measured by
$(p,n)$ reactions \protect\cite{krofcheck85,krofcheck87}.}
\begin{tabular}{lc}
\multicolumn{1}{c}{$E_{\rm ex}$}&${\rm BGT/(BGT)}_{\rm g.s.}$\\
\multicolumn{1}{c}{(MeV)}\\ \hline
0.0&1.000\\
0.175&\llap{$<$} 0.056\\
0.50&0.146\\
0.80&0.451\\
1.25&0.404\\
1.75&0.485\\
2.25&0.443\\
2.75&1.101\\
3.25&1.680\\
3.75&2.746\\
4.25&3.300\\
4.75&3.380\\
5.25&3.265\\
5.75&5.387\\
6.25&5.944\\
6.75&4.924\\
7.20&1.573
\label{BGT}
\end{tabular}
\end{minipage}
\end{table}

\begin{table}[htb]
\centering
\tighten
\caption[]{Best-Estimate Absorption Cross Sections for Specific
Energies.  The energies, $q$, are expressed in MeV and the cross
sections, $\sigma$, in units of $10^{-46}~{\rm
cm^2}$.\protect\label{bestsigmacue}}
\begin{tabular}{rrrrrr}
\noalign{\smallskip}
\multicolumn{1}{c}{$q$}&\multicolumn{1}{c}{$\sigma$}&\multicolumn{1}{c}{$q$}&\multicolumn{1}{c}{$\sigma$}&\multicolumn{1}{c}{$q$}&\multicolumn{1}{c}{$\sigma$}\\
\noalign{\smallskip}
\hline
\noalign{\smallskip}
 0.240& 1.310E+01& 1.500& 2.243E+02&10.000& 5.710E+04\\
 0.250& 1.357E+01& 1.600& 2.553E+02&10.500& 6.705E+04\\
 0.275& 1.499E+01& 1.700& 2.886E+02&11.000& 7.781E+04\\
 0.300& 1.662E+01& 1.750& 3.061E+02&11.500& 8.937E+04\\
 0.325& 1.836E+01& 2.000& 3.972E+02&12.000& 1.017E+05\\
 0.350& 2.018E+01& 2.500& 6.493E+02&12.500& 1.148E+05\\
 0.375& 2.208E+01& 3.000& 9.905E+02&13.000& 1.287E+05\\
 0.400& 2.406E+01& 3.500& 1.464E+03&13.500& 1.432E+05\\
 0.425& 2.648E+01& 4.000& 2.129E+03&14.000& 1.585E+05\\
 0.450& 2.862E+01& 4.500& 3.074E+03&14.500& 1.745E+05\\
 0.500& 3.314E+01& 5.000& 4.380E+03&15.000& 1.912E+05\\
 0.600& 4.300E+01& 5.500& 6.133E+03&15.500& 2.085E+05\\
 0.700& 5.397E+01& 6.000& 8.434E+03&16.000& 2.264E+05\\
 0.800& 6.848E+01& 6.500& 1.144E+04&18.000& 3.040E+05\\
 0.900& 8.276E+01& 7.000& 1.530E+04&20.000& 3.899E+05\\
 1.000& 9.830E+01& 7.500& 2.009E+04&22.500& 5.064E+05\\
 1.100& 1.226E+02& 8.000& 2.576E+04&25.000& 6.296E+05\\
 1.200& 1.440E+02& 8.500& 3.230E+04&30.000& 8.789E+05\\
 1.300& 1.672E+02& 9.000& 3.968E+04&\\
 1.400& 1.921E+02& 9.500& 4.797E+04\\
\noalign{\smallskip}
\end{tabular}
\end{table}

\begin{table}[htb]
\centering
\tighten
\caption[]{The $3\sigma$ Lower Limit Cross Sections.  The energies, $q$, are expressed in MeV and the cross
sections, $\sigma$, in units of $10^{-46}~{\rm
cm^2}$.\protect\label{minsigmacue}}
\begin{tabular}{rrrrrr}
\noalign{\smallskip}
\multicolumn{1}{c}{$q$}&\multicolumn{1}{c}{$\sigma$}&\multicolumn{1}{c}{$q$}&\multicolumn{1}{c}{$\sigma$}&\multicolumn{1}{c}{$q$}&\multicolumn{1}{c}{$\sigma$}\\
\noalign{\smallskip}
\hline
\noalign{\smallskip}
 0.240& 1.224E+01& 1.500& 1.882E+02&10.000& 3.040E+04\\
 0.250& 1.268E+01& 1.600& 2.122E+02&10.500& 3.541E+04\\
 0.275& 1.400E+01& 1.700& 2.378E+02&11.000& 4.082E+04\\
 0.300& 1.551E+01& 1.750& 2.511E+02&11.500& 4.657E+04\\
 0.325& 1.714E+01& 2.000& 3.162E+02&12.000& 5.264E+04\\
 0.350& 1.884E+01& 2.500& 4.971E+02&12.500& 5.902E+04\\
 0.375& 2.060E+01& 3.000& 7.318E+02&13.000& 6.568E+04\\
 0.400& 2.244E+01& 3.500& 1.041E+03&13.500& 7.260E+04\\
 0.425& 2.465E+01& 4.000& 1.456E+03&14.000& 7.973E+04\\
 0.450& 2.664E+01& 4.500& 2.017E+03&14.500& 8.706E+04\\
 0.500& 3.083E+01& 5.000& 2.766E+03&15.000& 9.455E+04\\
 0.600& 3.996E+01& 5.500& 3.744E+03&15.500& 1.022E+05\\
 0.700& 5.011E+01& 6.000& 5.001E+03&16.000& 1.098E+05\\
 0.800& 6.245E+01& 6.500& 6.615E+03&18.000& 1.406E+05\\
 0.900& 7.505E+01& 7.000& 8.658E+03&20.000& 1.689E+05\\
 1.000& 8.868E+01& 7.500& 1.117E+04&22.500& 1.955E+05\\
 1.100& 1.076E+02& 8.000& 1.412E+04&25.000& 2.056E+05\\
 1.200& 1.251E+02& 8.500& 1.752E+04&30.000& 1.484E+05\\
 1.300& 1.439E+02& 9.000& 2.136E+04\\
 1.400& 1.640E+02& 9.500& 2.566E+04\\
\noalign{\smallskip}
\end{tabular}
\end{table}

\begin{table}[htb]
\centering
\tighten
\caption[]{The $3\sigma$ Maximum Cross Sections.  The energies, $q$, are expressed in MeV and the cross
sections, $\sigma$, in units of $10^{-46}~{\rm
cm^2}$.\protect\label{maxsigmacue}}
\begin{tabular}{rrrrrr}
\noalign{\smallskip}
\multicolumn{1}{c}{$q$}&\multicolumn{1}{c}{$\sigma$}&\multicolumn{1}{c}{$q$}&\multicolumn{1}{c}{$\sigma$}&\multicolumn{1}{c}{$q$}&\multicolumn{1}{c}{$\sigma$}\\
\noalign{\smallskip}
\hline
\noalign{\smallskip}
 0.240& 1.395E+01& 1.500& 3.311E+02&10.000& 1.123E+05\\
 0.250& 1.446E+01& 1.600& 3.829E+02&10.500& 1.321E+05\\
 0.275& 1.598E+01& 1.700& 4.390E+02&11.000& 1.536E+05\\
 0.300& 1.772E+01& 1.750& 4.685E+02&11.500& 1.766E+05\\
 0.325& 1.958E+01& 2.000& 6.336E+02&12.000& 2.013E+05\\
 0.350& 2.153E+01& 2.500& 1.073E+03&12.500& 2.275E+05\\
 0.375& 2.357E+01& 3.000& 1.688E+03&13.000& 2.552E+05\\
 0.400& 2.567E+01& 3.500& 2.559E+03&13.500& 2.845E+05\\
 0.425& 3.101E+01& 4.000& 3.808E+03&14.000& 3.152E+05\\
 0.450& 3.362E+01& 4.500& 5.609E+03&14.500& 3.474E+05\\
 0.500& 3.921E+01& 5.000& 8.127E+03&15.000& 3.811E+05\\
 0.600& 5.154E+01& 5.500& 1.153E+04&15.500& 4.161E+05\\
 0.700& 6.534E+01& 6.000& 1.603E+04&16.000& 4.526E+05\\
 0.800& 8.261E+01& 6.500& 2.194E+04&18.000& 6.117E+05\\
 0.900& 1.031E+02& 7.000& 2.954E+04&20.000& 7.910E+05\\
 1.000& 1.258E+02& 7.500& 3.900E+04&22.500& 1.041E+06\\
 1.100& 1.658E+02& 8.000& 5.022E+04&25.000& 1.317E+06\\
 1.200& 1.996E+02& 8.500& 6.316E+04&30.000& 1.930E+06\\
 1.300& 2.366E+02& 9.000& 7.779E+04\\
 1.400& 2.768E+02& 9.500& 9.422E+04\\
\noalign{\smallskip}
\end{tabular}
\end{table}

\begin{table}[htb]
\centering
\tighten
\begin{minipage}[htb]{5in}
\caption[]{The Minimum Solar Neutrino Rates if Selected Nuclear
Reactions Are Set Equal to Zero.  The
$1\sigma$ uncertainty in the chlorine rate is about $\pm 0.01$ SNU in
all three cases.\protect\label{minrates}}
\begin{tabular}{lcc}
\noalign{\smallskip}
\multicolumn{1}{c}{Reactions Set Equal to Zero}&Ga Rate&Cl Rate\\
&(SNU)&(SNU)\\
\noalign{\smallskip}
\hline
\noalign{\smallskip}
\multicolumn{1}{c}{${\rm ^3He}(\alpha,\gamma){\rm ^7Be}$}&$88.1^{+3.2}_{-2.4}$&0.7\\
\noalign{\smallskip}
${\rm ^3He}(\alpha,\gamma){\rm ^7Be}$\ ,\ \ ${\rm ^{12}C}(p, \gamma){\rm ^{13}N}$&$79.7^{+2.4}_{-2.0}$&0.3\\
\noalign{\smallskip}
${\rm ^3He}(\alpha, \gamma){\rm ^7Be}$\ ,\ \ all CNO reactions&$79.5^{+2.3}_{-2.0}$&0.3\\
\noalign{\smallskip}
\end{tabular}
\end{minipage}
\end{table}

\begin{table}[htb]
\centering
\tighten
\caption[]{Neutrino Absorption Cross Sections for Standard Energy
spectra.  All cross sections are given in units of $10^{-46}\ {\rm
cm}^2$ except for ${\rm ^8B}$ and $hep$ neutrinos, for which the unit is
$10^{-42}\ {\rm cm}^2$.  The uncertainties indicated are effective
$1\sigma$ uncertainties.\protect\label{crosssummary}}
\begin{tabular}{lcccccccccc}
\noalign{\smallskip}
\multicolumn{1}{c}{Source}&$pp$&$pep$&$hep$&${\rm ^7Be}$&${\rm ^8B}$&${\rm
^{13}N}$&${\rm ^{15}O}$&${\rm ^{17}F}$&${\rm ^{37}Ar}$&${\rm ^{51}Cr}$\\
\noalign{\smallskip}
\hline
Best&11.72&204&7.14&71.7&2.40&60.4&113.7&113.9&70.0&58.1\\
$1\sigma_+ (\%)$&2.3&17&32&7&32&6&12&12&0.07&0.04\\
$1\sigma_- (\%)$&2.3&7&16&3&15&3&6&6&0.03&0.03\\
\noalign{\smallskip}
\end{tabular}
\end{table}

\begin{table}[htb]
\centering
\tighten
\caption[]{The $pp$ neutrino spectrum. 
The normalized $pp$ neutrino energy
spectrum, $P$($q$), 
is given in intervals of 5.0406 keV.  The neutrino energy, $q$, is expressed
in MeV and $P$($q$) is normalized per MeV.\protect\label{ppnumerical}}
\begin{tabular}{lccccccc}
\noalign{\smallskip}
\multicolumn{1}{c}{$q$}&$P(q)$&$q$&$P(q)$&$q$&$P(q)$
&$q$&$P(q)$\\
\noalign{\smallskip}
\hline
\noalign{\smallskip}
0.00504&  0.0035&  0.11089&  1.2477&  0.21675&  3.2300&  0.32260&  4.0356\\
0.01008&  0.0138&  0.11593&  1.3417&  0.22179&  3.3094&  0.32764&  4.0114\\
0.01512&  0.0307&  0.12097&  1.4370&  0.22683&  3.3859&  0.33268&  3.9794\\
0.02016&  0.0538&  0.12601&  1.5335&  0.23187&  3.4594&  0.33772&  3.9391\\
0.02520&  0.0830&  0.13106&  1.6310&  0.23691&  3.5298&  0.34276&  3.8900\\
0.03024&  0.1179&  0.13610&  1.7291&  0.24195&  3.5966&  0.34780&  3.8316\\
0.03528&  0.1582&  0.14114&  1.8278&  0.24699&  3.6599&  0.35284&  3.7632\\
0.04032&  0.2038&  0.14618&  1.9267&  0.25203&  3.7194&  0.35788&  3.6842\\
0.04537&  0.2543&  0.15122&  2.0258&  0.25707&  3.7749&  0.36292&  3.5937\\
0.05041&  0.3094&  0.15626&  2.1247&  0.26211&  3.8262&  0.36796&  3.4907\\
0.05545&  0.3691&  0.16130&  2.2233&  0.26715&  3.8731&  0.37300&  3.3740\\
0.06049&  0.4329&  0.16634&  2.3214&  0.27219&  3.9154&  0.37804&  3.2422\\
0.06553&  0.5006&  0.17138&  2.4187&  0.27723&  3.9529&  0.38309&  3.0932\\
0.07057&  0.5721&  0.17642&  2.5151&  0.28227&  3.9854&  0.38813&  2.9246\\
0.07561&  0.6469&  0.18146&  2.6105&  0.28731&  4.0127&  0.39317&  2.7330\\
0.08065&  0.7250&  0.18650&  2.7044&  0.29235&  4.0344&  0.39821&  2.5136\\
0.08569&  0.8061&  0.19154&  2.7969&  0.29740&  4.0505&  0.40325&  2.2589\\
0.09073&  0.8899&  0.19658&  2.8877&  0.30244&  4.0605&  0.40829&  1.9567\\
0.09577&  0.9761&  0.20162&  2.9766&  0.30748&  4.0644&  0.41333&  1.5832\\
0.10081&  1.0647&  0.20666&  3.0634&  0.31252&  4.0617&  0.41837&  1.0783\\
0.10585&  1.1553&  0.21171&  3.1479&  0.31756&  4.0522&  0.42341&  0.0000\\
\noalign{\smallskip}
\end{tabular}
\end{table}

\begin{table}[htb]
\centering
\tighten
\caption[]{The ${\rm ^{13}N}$ neutrino spectrum. 
The normalized ${\rm ^{13}N}$ neutrino energy
spectrum, $P$($q$), 
is given in intervals of 14.264 keV.  The neutrino energy, $q$, is expressed
in MeV and $P$($q$) is normalized per MeV.\protect\label{13Nnumerical}}
\begin{tabular}{lccccccc}
\noalign{\smallskip}
\multicolumn{1}{c}{$q$}&$P(q)$&$q$&$P(q)$&$q$&$P(q)$
&$q$&$P(q)$\\
\noalign{\smallskip}
\hline
\noalign{\smallskip}
0.01426&0.0018&0.31381&0.5787&0.61336&1.2865&0.91291&1.2732\\
0.02853&0.0071&0.32808&0.6185&0.62763&1.3066&0.92718&1.2472\\
0.04279&0.0157&0.34234&0.6583&0.64189&1.3248&0.94144&1.2186\\
0.05706&0.0275&0.35661&0.6981&0.65616&1.3413&0.95571&1.1875\\
0.07132&0.0422&0.37087&0.7378&0.67042&1.3558&0.96997&1.1538\\
0.08559&0.0596&0.38514&0.7771&0.68469&1.3684&0.98424&1.1175\\
0.09985&0.0796&0.39940&0.8160&0.69895&1.3790&0.99850&1.0784\\
0.11411&0.1020&0.41366&0.8544&0.71321&1.3876&1.01276&1.0366\\
0.12838&0.1267&0.42793&0.8922&0.72748&1.3940&1.02703&0.9918\\
0.14264&0.1534&0.44219&0.9293&0.74174&1.3983&1.04129&0.9440\\
0.15691&0.1820&0.45646&0.9656&0.75601&1.4005&1.05556&0.8931\\
0.17117&0.2124&0.47072&1.0010&0.77027&1.4005&1.06982&0.8387\\
0.18544&0.2444&0.48499&1.0353&0.78454&1.3983&1.08409&0.7805\\
0.19970&0.2777&0.49925&1.0686&0.79880&1.3937&1.09835&0.7183\\
0.21396&0.3124&0.51351&1.1008&0.81306&1.3870&1.11261&0.6512\\
0.22823&0.3482&0.52778&1.1316&0.82733&1.3778&1.12688&0.5785\\
0.24249&0.3849&0.54204&1.1612&0.84159&1.3664&1.14114&0.4987\\
0.25676&0.4226&0.55631&1.1893&0.85586&1.3526&1.15541&0.4094\\
0.27102&0.4609&0.57057&1.2160&0.87012&1.3364&1.16967&0.3057\\
0.28529&0.4997&0.58484&1.2411&0.88439&1.3178&1.18394&0.1766\\
0.29955&0.5391&0.59910&1.2646&0.89865&1.2967&1.19820&0.0000\\
\noalign{\smallskip}
\end{tabular}
\end{table}

\begin{table}[htb]
\centering
\tighten
\caption[]{The ${\rm ^{15}O}$ neutrino spectrum. 
The normalized ${\rm ^{15}O}$ neutrino energy
spectrum, $P$($q$), 
is given in intervals of 20.615 keV.  The neutrino energy, $q$, is expressed
in MeV and $P$($q$) is normalized per MeV.\protect\label{15Onumerical}}
\begin{tabular}{lccccccc}
\noalign{\smallskip}
\multicolumn{1}{c}{$q$}&$P(q)$&$q$&$P(q)$&$q$&$P(q)$
&$q$&$P(q)$\\
\noalign{\smallskip}
\hline
\noalign{\smallskip}
0.02062&0.0014&0.45354&0.4372&0.88647&0.9227&1.31939&0.8354\\
0.04123&0.0056&0.47416&0.4663&0.90708&0.9341&1.34001&0.8136\\
0.06185&0.0123&0.49477&0.4954&0.92770&0.9441&1.36062&0.7902\\
0.08246&0.0214&0.51539&0.5242&0.94831&0.9527&1.38124&0.7652\\
0.10308&0.0328&0.53600&0.5528&0.96893&0.9597&1.40185&0.7386\\
0.12369&0.0463&0.55662&0.5810&0.98954&0.9652&1.42247&0.7106\\
0.14431&0.0617&0.57723&0.6088&1.01016&0.9692&1.44308&0.6809\\
0.16492&0.0790&0.59785&0.6360&1.03077&0.9716&1.46370&0.6497\\
0.18554&0.0979&0.61846&0.6626&1.05139&0.9725&1.48431&0.6170\\
0.20615&0.1184&0.63908&0.6886&1.07200&0.9717&1.50493&0.5827\\
0.22677&0.1402&0.65970&0.7138&1.09262&0.9693&1.52555&0.5467\\
0.24739&0.1634&0.68031&0.7381&1.11324&0.9653&1.54616&0.5091\\
0.26800&0.1876&0.70093&0.7616&1.13385&0.9596&1.56678&0.4697\\
0.28862&0.2129&0.72154&0.7840&1.15447&0.9523&1.58739&0.4283\\
0.30923&0.2391&0.74216&0.8055&1.17508&0.9434&1.60801&0.3846\\
0.32985&0.2660&0.76277&0.8259&1.19570&0.9329&1.62862&0.3384\\
0.35046&0.2936&0.78339&0.8451&1.21631&0.9207&1.64924&0.2888\\
0.37108&0.3217&0.80400&0.8632&1.23693&0.9069&1.66985&0.2348\\
0.39169&0.3502&0.82462&0.8800&1.25754&0.8914&1.69047&0.1737\\
0.41231&0.3790&0.84523&0.8956&1.27816&0.8744&1.71108&0.0998\\
0.43293&0.4080&0.86585&0.9098&1.29877&0.8557&1.73170&0.0000\\
\noalign{\smallskip}
\end{tabular}
\end{table}

\begin{table}[htb]
\centering
\tighten
\caption[]{The ${\rm ^{17}F}$ neutrino spectrum. 
The normalized ${\rm ^{17}F}$ neutrino energy
spectrum, $P$($q$), 
is given in intervals of 20.671 keV.  The neutrino energy, $q$, is expressed
in MeV and $P$($q$) is normalized per MeV.\protect\label{17Fnumerical}}
\begin{tabular}{lccccccc}
\noalign{\smallskip}
\multicolumn{1}{c}{$q$}&$P(q)$&$q$&$P(q)$&$q$&$P(q)$
&$q$&$P(q)$\\
\noalign{\smallskip}
\hline
\noalign{\smallskip}
0.02067&0.0014&0.45477&0.4375&0.88887&0.9223&1.32297&0.8323\\
0.04134&0.0056&0.47544&0.4666&0.90954&0.9337&1.34364&0.8103\\
0.06201&0.0123&0.49611&0.4957&0.93021&0.9436&1.36431&0.7866\\
0.08269&0.0214&0.51679&0.5245&0.95089&0.9521&1.38499&0.7614\\
0.10336&0.0328&0.53746&0.5531&0.97156&0.9590&1.40566&0.7347\\
0.12403&0.0463&0.55813&0.5813&0.99223&0.9645&1.42633&0.7064\\
0.14470&0.0618&0.57880&0.6091&1.01290&0.9683&1.44700&0.6765\\
0.16537&0.0791&0.59947&0.6363&1.03357&0.9706&1.46767&0.6451\\
0.18604&0.0980&0.62014&0.6629&1.05424&0.9713&1.48834&0.6121\\
0.20671&0.1185&0.64081&0.6888&1.07491&0.9704&1.50901&0.5776\\
0.22739&0.1404&0.66149&0.7140&1.09559&0.9679&1.52969&0.5414\\
0.24806&0.1635&0.68216&0.7383&1.11626&0.9638&1.55036&0.5035\\
0.26873&0.1878&0.70283&0.7617&1.13693&0.9580&1.57103&0.4638\\
0.28940&0.2131&0.72350&0.7841&1.15760&0.9506&1.59170&0.4221\\
0.31007&0.2393&0.74417&0.8056&1.17827&0.9415&1.61237&0.3782\\
0.33074&0.2662&0.76484&0.8259&1.19894&0.9308&1.63304&0.3317\\
0.35141&0.2938&0.78551&0.8451&1.21961&0.9184&1.65371&0.2818\\
0.37209&0.3219&0.80619&0.8631&1.24029&0.9044&1.67439&0.2275\\
0.39276&0.3504&0.82686&0.8799&1.26096&0.8888&1.69506&0.1663\\
0.41343&0.3793&0.84753&0.8954&1.28163&0.8716&1.71573&0.0927\\
0.43410&0.4083&0.86820&0.9095&1.30230&0.8527&1.73640&0.0000\\
\noalign{\smallskip}
\end{tabular}
\end{table}

\newpage
\begin{figure}[htb]
\centerline{\psfig{figure=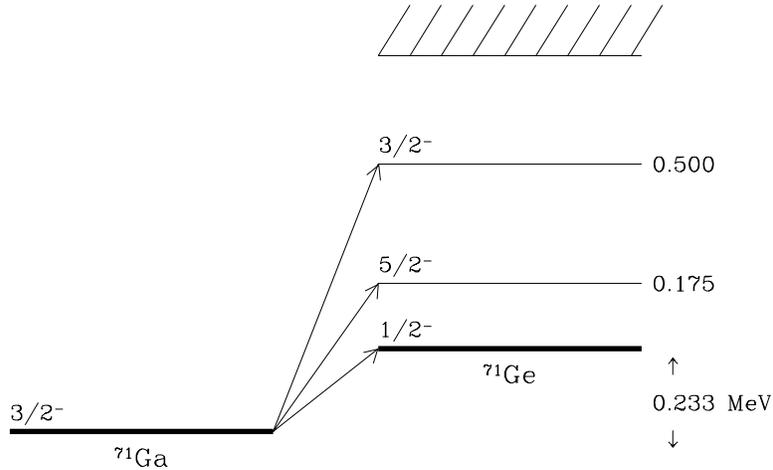,width=4in}}
\caption[]{The ${\rm ^{71}Ga}$-${\rm ^{71}Ge}$ Transitions for Low
Energy Neutrinos.  Only the
ground state and the first two allowed excited state transitions
contribute to the absorption of $pp$, ${\rm ^7Be}$, and ${\rm
^{51}Cr}$ neutrinos.  The ${\rm ^8B}$, CNO, and $pep$ neutrinos all
give rise to excited state transitions that are unconstrained by the
${\rm ^{51}Cr}$ neutrino absorption measurements and for which the
$(p,n)$ measurements provide the only empirical guide to the relevant
BGT values.\protect\label{gagefigure}}
\end{figure}

\begin{figure}[htb]
\centerline{\psfig{figure=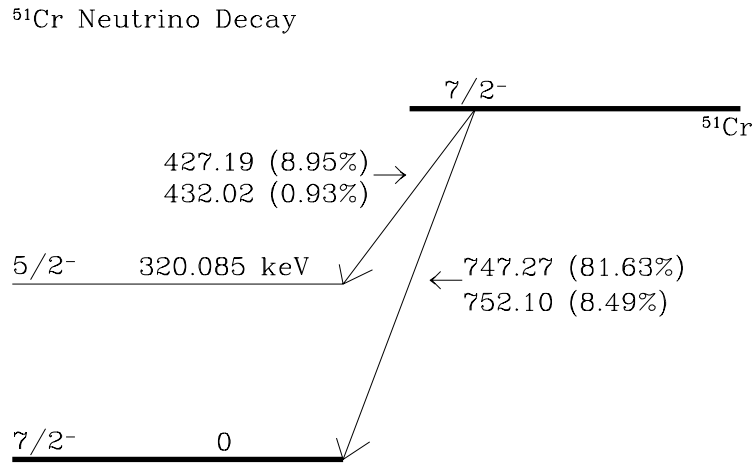,width=4in}}
\caption[]{The ${\rm ^{51}Cr}$ Decay Scheme.\label{crfigure}}
\label{crdecayscheme}
\end{figure}

\begin{figure}[htb]
\centerline{\psfig{figure=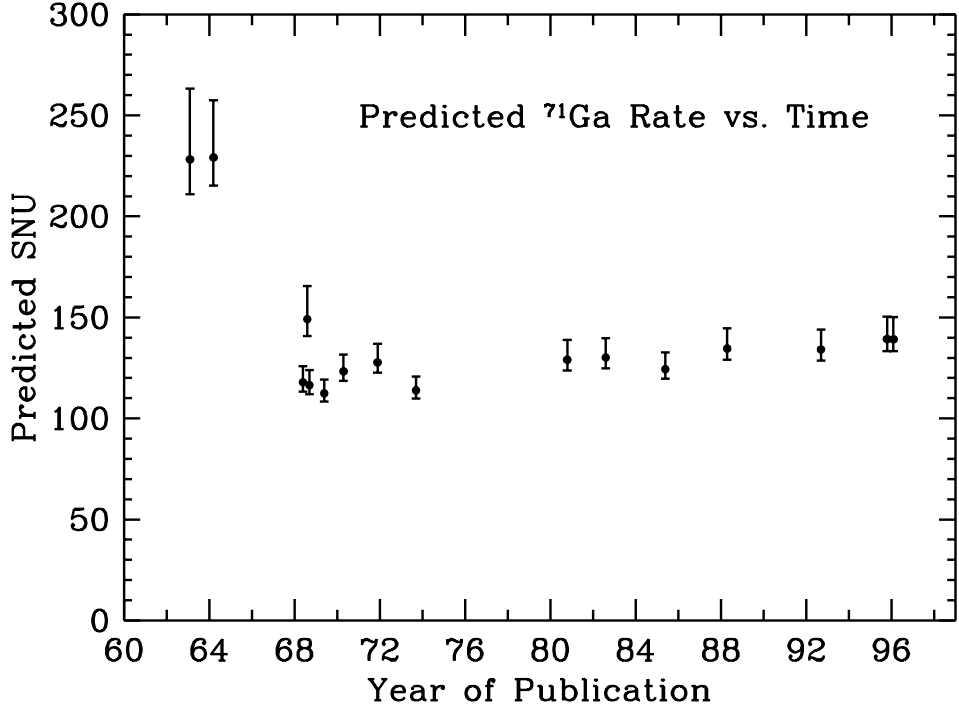,width=5in}}
\vglue.2in
\caption[]{ Predicted Solar Neutrino Gallium Event Rate Versus 
Year of Publication.  The figure shows the event rates for all of the
standard 
solar model calculations that my colleagues and I have
published\cite{BP92,BP95,fibs,bahcall97,solarpapers}.  
The cross 
sections  from the present paper 
have been used in all cases to convert the calculated neutrino fluxes to 
predicted capture rates.  
The estimated $1\sigma$ uncertainties reflect in all cases just the 
uncertainties in the cross sections that are evaluated in the present paper.
For the 35 years over which we have been calculating standard
solar model neutrino fluxes, the historically lowest value (fluxes 
published in 1969)
corresponds to $109.5$~SNU.  
This lowest-ever value is $5.6\sigma$ greater than the
combined GALLEX and SAGE experimental result.
If the points prior to 1992 are increased by 
\hbox{$11$~SNU} to correct for diffusion (this was not done in the
figure), 
then all of the standard model 
theoretical capture rates since 1968 through 1997 lie in the range 
$120$ SNU to $141$ SNU, i.e., ($131 \pm 11$) SNU. 
\protect\label{gahistory}}
\end{figure} 

\begin{figure}[htb]
\centerline{\psfig{figure=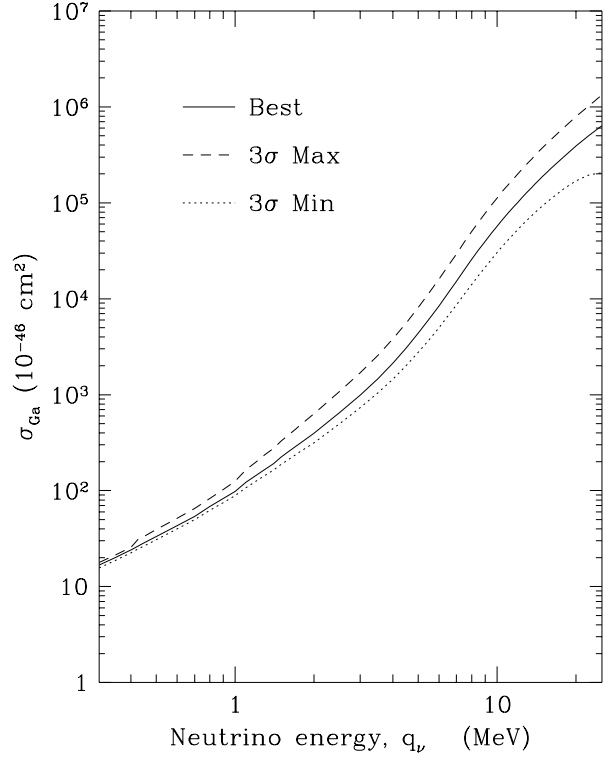,width=4in}}
\caption[]{ Absorption cross sections 
for gallium as a function of energy.  The figure displays
the best-estimate cross sections as well as the $\pm 3\sigma$ 
cross sections.  Numerical values are given in Table~\ref{bestsigmacue},
Table~\ref{minsigmacue}, and Table~\ref{maxsigmacue}\protect\label{sigmaque}.
}
\end{figure}

\end{document}